\documentclass[11pt]{JHEP}
\usepackage{latexsym}
\usepackage{youngtab}
\usepackage{amssymb,amsfonts}
\usepackage{amsmath,amsthm,epsfig,euscript,array,cancel}
\font\mybb=msbm10 at 11pt
\def\bb#1{\hbox{\mybb#1}}

\preprint

\preprint{hep-th/1612.01321
 V3: August 22, 2017}

\title{Exceptional field theories,
 superparticles in an enlarged 11D superspace and higher spin theories.}

\author{
Igor Bandos
\\ \vspace{0.8cm}
$^\dagger$ Department of
Theoretical Physics, University of the Basque Country UPV/EHU,
\\ P.O. Box 644, 48080 Bilbao, Spain
 \\
 $^\ddagger$ IKERBASQUE, Basque Foundation for Science, 48011, Bilbao, Spain
 }

\date{January-- December 2016}

\abstract{Recently proposed exceptional field theories (EFTs) making manifest the duality $E_{n(n)}$ symmetry, first observed as nonlinearly realized symmetries of the maximal $d=3,4,...,9$ supergravity ($n=11-d$) and containing 11D and type IIB supergravity as sectors,  were formulated in enlarged  spacetimes. In the case of $E_{7(7)}$ EFT such an  enlarged spacetime can be identified with the bosonic body of the $d=4$ central charge superspace $\Sigma^{(60|32)}$, the ${\cal N}=8$ $d=4$  superspace completed by  56 additional bosonic coordinates associated to central charges of the maximal $d=4$ supersymmetry algebra.

In this paper we show how the hypothesis on the relation of all the known $E_{n(n)}$ EFTs, including $n=8$, with supersymmetry leads  to the conjecture on  existence of 11D exceptional field theory  living in  11D tensorial central charge superspace $\Sigma^{(528|32)}$ and underlying all the $E_{n(n)}$ EFTs with $n=2,...,8$, and probably the double field theory (DFT). We conjecture the possible form of the section conditions of such an 11D EFT and show that quite generic solutions of these can be generated by superparticle models the ground states of which preserve from one half to  all but one supersymmetry. The properties of these superparticle models are briefly discussed. We argue that, upon quantization,  their quantum states should describe  free massless non-conformal higher spin fields in D=11.
We also discuss some relevant representations of the M-theory superalgebra which, in the present context, describes  supersymmetry of the 11D EFT.
 }

\keywords{Supersymmetry, U-duality, superspace, superparticle, higher spin theory, double field theory, exceptional field theories }
\date{05/12/2016,
V2: 19/12/16,
V2': 24/01/2017, V3: 20/08/2017}

\begin{document}

\maketitle

\setcounter{page}2

\newpage

\section{Introduction}

Recently exceptional field theories (EFTs) \cite{Hohm:2013pua}, manifestly invariant under U--duality symmetry groups
$E_{n(n)}$ \cite{Hull:1994ys}  with $n=2,3,4,5,6,7,8$ and containing 11D and 10D type IIB supergravity theories as sectors were formulated in enlarged d=3,4,5,...,9 spaces \cite{Hull:2007zu,Berman:2011cg,Berman:2012vc,Coimbra:2012af,Hohm:2013pua,Hohm:2013vpa,Hohm:2013uia,Hohm:2014fxa,Godazgar:2014nqa,
Musaev:2014lna,Ciceri:2014wya,Hohm:2015xna,Abzalov:2015ega,Cederwall:2015ica,Musaev:2015ces,Berman:2015rcc,
Cederwall:2016ukd,Baguet:2016jph} \footnote{The embedding of massive IIA requires a deformation of (section conditions of) the EFT \cite{Ciceri:2016dmd,Cassani:2016ncu}.}. The value of  $d$ is related to $n$ by $d+n=11$, and in this sense one can call the $E_{n(n)}$ EFT '$d$-dimensional', the name which also reflects its manifest invariance under the $d$-dimensional Lorentz group  $SO(1,d-1)$
\footnote{ It is also worth commenting that, while $E_{6(6)}$, $E_{7(7)}$ and $E_{8(8)}$ of EFTs with manifest d=5,4,3 Lorentz symmetries are the exceptional Lie groups from the Cartan list, for lower $n$ $E_{n(n)}$ denote simpler groups:  $E_{5(5)}=SO(5,5)$, $E_{4(4)}=SL(5)$, $E_{3(3)}=SL(2)\times SL(2)$ and, as it was proposed in recent \cite{Berman:2015rcc}, $E_{2(2)}=SL(2)\times {\bb R}^+$.}.   They can be regarded as M-theoretic counterparts of D=10 double field theory (DFT) \cite{Siegel:1993xq,Hull:2004in,Hull:2009mi,Coimbra:2011nw,Jeon:2011vx,Hohm:2013bwa,Park:2013mpa,
Hatsuda:2014qqa,Hatsuda:2015cia}  designed to have a manifest T-duality symmetry, characteristic for string theory \footnote{See \cite{Green:1987sp} and refs. therein for T-duality and
\cite{Tseytlin:1990nb,Berman:2007xn,Copland:2011wx,Blair:2013noa,DeAngelis:2013wba,Hatsuda:2015cia,Bandos:2015cha,
Blair:2016xnn,Park:2016sbw}
for string and superstring in doubled (super)spaces. Notice also that we usually denote the number of spacetime dimensions by $D$ when it is equal to 10 or 11, and by $d$ when it is lower, so that $d\leq 9$.}. The DFT is formulated in the space with doubled number, 2D, of bosonic coordinates (usually D=10 is assumed in this case).
The number of the additional bosonic coordinates $y^\Sigma$ of the $d$ dimensional $E_{n(n)}$ EFT is d-/n- dependent: it varies from 3 in the recently proposed 9d `F-theory action' of \cite{Berman:2015rcc}
to 56 in d=4  $E_{7(7)}$ EFT \cite{Hohm:2013pua,Hohm:2013uia,Godazgar:2014nqa}
and 248 in $d=3$ $E_{8(8)}$ EFT \cite{Hohm:2014fxa,Cederwall:2015ica}. The dependence of the fields on additional coordinates is restricted by the so--called {\it section conditions}\footnote{The name `section conditions' was  introduced in \cite{Berman:2011cg} developing $E_{4(4)}=SL(5)$ (pre-)EFT formalism of \cite{Berman:2010is}. The name EFT was introduced in  \cite{Hohm:2013pua} which starts a series of papers formulating the EFTs for  exceptional U-duality groups $E_{7,7}$, $E_{6(6)}$ and $E_{8(8)}$  in its complete form, including all the differential form fields of maximal $d=11-n$ dimensional supergravity.}
 the strong version of which is imposed (`by hand') on any pair of functions of the theory.

\begin{center}
    \begin{tabular}{ | l | l | l | l | p{5cm} |}
    \hline
    $E_{n(n)}$ & n & d=11-n & $N_n$= $\#$ of $y^\Sigma$   & Section condition and ref. \\ \hline  & & & &  \\
    $E_{8(8)}$ & 8 & d=3 & 248   & $Y_{\Lambda\Xi}{}^{{\Sigma}{\Pi}}\partial_{{\Sigma}} \otimes \partial_{{\Pi}} =0,\quad $ see \cite{Hohm:2014fxa}\\   $E_{7(7)}$ & 7 & d=4 & 56  & $t_{G}^{{\Sigma}{\Pi}}\partial_{{\Sigma}} \otimes \partial_{{\Pi}} =0$, \cite{Hohm:2013uia}, see below \\
    $E_{6(6)}$ & 6 & d=5 & 27  &  $d^{{\Lambda}{\Sigma}{\Pi}}\partial_{{\Sigma}} \otimes \partial_{{\Pi}} =0$  \cite{Hohm:2013pua}  \\
    $E_{5(5)}=SO(5,5)$ & 5 & d=6 & 16  & $\gamma_{I}^{{\Sigma}{\Pi}}\partial_{{\Sigma}} \otimes \partial_{{\Pi}} =0$ \cite{Abzalov:2015ega} \\
    $E_{4(4)}=SL(5)$ & 4 & d=7 & 10 ($y^{\tilde{\mathfrak{a}}\tilde{\mathfrak{b}}}=y^{[\tilde{\mathfrak{a}}\tilde{\mathfrak{b}}]}$) & $\partial_{[\tilde{\mathfrak{a}}\tilde{\mathfrak{b}}}\otimes \partial_{\tilde{\mathfrak{c}}\tilde{\mathfrak{d}}]}=0$ \cite{Berman:2011cg,Blair:2013gqa}, see below \\
    $E_{3(3)}=SL(3)\times SL(2)$ & 3 & d=8 & 6 $(y^{\alpha i})$  & $\epsilon^{ijk}\epsilon^{\alpha\beta} \partial_{{\alpha} i} \otimes \partial_{{\beta} j} =0$  \cite{Hohm:2015xna}
    \\ $E_{2(2)}=SL(2)\times {\bb R}^+$ & 2 & d=9 & 3 $(y^\alpha, z)$ & $ \partial_{z} \otimes \partial_{\alpha}+ \partial_{\alpha} \otimes \partial_{z}  =0$ \cite{Berman:2015rcc} \\ &&&& \\
     \hline
    \end{tabular}

\medskip

Table 1. {\small\it Additional coordinates and section conditions of the $E_{n(n)}$ EFTs. The notation for $n=7$ and $n=4$ cases are described below. The other cases will not be discussed and we refer to the  original papers
(cited at the end of the lines) for the notation.\footnote{Notice a partial intersection of (the 'left hand side' of) this Table 1 with Table 2 of \cite{Duff:1990hn}, where a possible relation of 11D supermembrane duality transformations with   $E_{n(n)}$ duality symmetries of dimensionally reduced maximal supergravity was discussed.} }
\end{center}

In the case of DFT the solution of the  strong  section conditions implies that all the physical fields depend only on D of 2D bosonic coordinates. The manifest T-duality is provided by the freedom in choosing the set of these $D$ of the complete set of $2D$ coordinates. This is called `choice of the section' (hence the name `section conditions' for the equations solved by this choice).

The structure of the EFT section conditions looks strongly d- (or n-)dependent and much less transparent.
As we will discuss below, the analysis of differences in the structure of EFTs with different $n$ suggests the possible existence (and makes desirable to find) a hypothetical {\it underlying EFT}, which we call '11D EFT' or `uEFT', such that  all the lower d EFTs can be obtained by its reductions. \footnote{Notice that our hypothetical uEFT is not identical but probably complementay to the E11 program of \cite{West:2001as,West:2003fc,West:2007mh,West:2010rv,Henneaux:2010ys,West:2014eza,Tumanov:2015iea,Tumanov:2015yjd,Tumanov:2016abm,Tumanov:2016dxc}. We will comment on this more in concluding Sec. 5. A discussion on  the connection of  EFTs and E11 can be found in  \cite{Tumanov:2015iea}. }

In this paper we make same stages toward the construction of such a hypothetical 11D uEFT.
In particular, we argue that the natural basis for its construction is provided by 11D tensorial central charge superspace $\Sigma^{(528|32)}$, propose  the section conditions for this uEFT in this superspace, and present a family of superparticle models in $\Sigma^{(528|32)}$ which produce quite generic solutions of these section conditions.  The quantum states of these models are massless, which allows to conjecture that their quantization  results in supersymmetric theories of free massless higher spin fields in D=11. The  quantization of D=10 version(s) of the model(s) should produce  a theory of free massless non-conformal higher spin field, the tower of which includes 10D 'graviton'.

To gain a hint about how the  quantum state spectrum of some of such models might look like, and also as an additional argument  in favour of our uEFT hypothesis, we discuss some unitary highest weight representations of M-theory superalgebra, which in our context describes supersymmetry of $\Sigma^{(528|32)}$ and of the hypothetical uEFT, and the embedding of 11D supergravity in these  representations.

The rest of this paper is organized as follows. In the beginning of next Sec. 2
 we review the structure of $E_{n(n)}$ exceptional field theories (EFTs) with $n=2,...,8$ and  conjecture on their relation with the most general supersymmetry algebra. In particular, in sec. 2.1 we discuss the additional coordinates of
$E_{n(n)}$ EFTs, the section conditions, which are imposed to restrict the dependence of EFT fields on those,  and their classical counterparts. In sec. 2.2 we argue in favor of relation of additional coordinates of $E_{n(n)}$ EFTs  with  maximal supersymmetry algebra in $d=11-n$, describe the relation  of those with  central charges of such a supersymmetry algebra observed first for  $n=7$. In sec. 2.3 we discuss the extension of this conjecture to  $n=8$ which requires involvement of also the vectorial 'central charges' and leads us to the most general d=3 maximal supersymmetry algebra.

The underlying EFT (11D EFT or uEFT) conjecture is formulated in Sec. 3.
In sec. 3.1 we show that the maximally extended d=3 supersymmetry  superalgebra has actually a bigger automorphism symmetry,  including  $SO(1,10)$, which allows us to call it M-algebra (or M-theory superalgebra),  describe  the $SO(1,10)$ invariant Cartan forms on the associated supergroup manifold  $\Sigma^{(528|32)}$ with 528 bosonic and 32 fermionic directions, and  conjecture on the existence of underlying 11D EFT, leaving in this superspace.

In Sec 3.2 we propose  the candidate section condition of 11D EFT needed to reduce the huge number of additional  bosonic coordinates and discuss the structure of their solutions. In sec. 3.3 we consider a series of superparticle models in $\Sigma^{(528|32)}$ which produce quite generic solution of the classical section conditions as their constraints. The actions of these models involve essentially 11D spinor moving frame variables \cite{Bandos:2007mi,Bandos:2007wm} (see also \cite{Bandos:1992np,Bandos:1992ze,
Bandos:1993yc,Bandos:2012hp}), also called Lorentz harmonics \cite{Galperin:1992pz} (see also
\cite{Sokatchev:1985tc,Sokatchev:1987nk,Bandos:1990ji,Galperin:1991gk,Delduc:1991ir,Bandos:1994eu,Uvarov:2000wt}); we describe these in sec. 3.3.3.

In sec. 4 we argue that the quantization of these uEFT-related superparticle models  should produce towers of massless 11D higher spin fields as their quantum state spectrum. We briefly describe (in secs. 4.1 and 4.2) the relation of lower dimensional  counterpart of preonic superparticle to free massless conformal higher spin fields in $D=4,6, 10$ dimensions and on this basis conjecture (in sec. 4.3) that the counterpart of  above mentioned generalized superparticle models with spinor moving frame variables provide the classical mechanic description of massless non-conformal higher spin fields in $D=6, 10$. The quantization of the models in $\Sigma^{(528|32)}$ should result in  a tower of massless non-conformal 11D higher spin fields; the conformal higher spin tower is not known for this case.

In sec. 5  we discuss unitary highest weight representations of M-theory superalgebra, which are relevant in the uEFT context, and the embedding of 11D supergravity in some of these  representations. We conclude in sec. 6 where the discussion on possible relation/complimentarity of our conjectured  11D EFT and of $E_{11}$ and $E_{10}$ hypothesis can be also found.

\section{On EFTs, their section conditions and  central charge superspaces}

\subsection{$E_{n(n)}$ EFTs with $n=2,...,8$. Additional coordinates, section conditions and  their classical counterparts}

Schematically, the $E_{n(n)}$ EFT is constructed on the basis of maximal $d=11-n$ dimensional supergravity (SUGRA) by allowing the field to depend, besides $d$ spacetime coordinates $x^\mu$, on additional 'internal' coordinates $y^\Sigma$ the number of which (i.e. the range of the index $\Sigma$), is given by dimension $N_n$ of minimal irreducible representation of $E_{n(n)}$ ($N_2=3$, $\ldots$, $N_7=56$, $N_8=248$). Besides that, the field strengths and the Lagrangian of $d$-dimensional supergravity  are modified by inclusion of terms with derivatives $\partial_\Sigma = \frac \partial {\partial y^\Sigma}$, and the action is constructed by integrating this modified SUGRA Lagrangian $ L^{^{E_{n,n}}}_{_{EFT}}$ over $d$ spacetime and all the $N_n$ internal coordinates, $S^{^{E_{n,n}}}_{_{EFT}}=\int d^{4}xd^{N_n}y\, L^{^{E_{n,n}}}_{_{EFT}}$. This integral is usually considered as formal as far as its rigid definition meets problems related with the next ingredients of EFT which we are going to describe now.

All the fields in EFT, $F$, are subject to the so-called weak section conditions
\begin{eqnarray}\label{sec=YdxdF=0}
& Y_{\Lambda\Xi}{}^{{\Sigma}{\Pi}}\partial_{{\Sigma}} \, \partial_{{\Pi}}F =0 \; , \qquad
\end{eqnarray}
where $Y_{\Lambda\Xi}{}^{{\Sigma}{\Pi}}$ is an invariant tensor of $E_{n(n)}$ the explicit  form of which  is strongly $n$-dependent. But moreover,
 all the {\it pairs} of the fields $F_1$, $F_2$,  should be subject to the so-called strong section conditions,
\begin{eqnarray}\label{sec=YdFdF=0}
& Y_{\Lambda\Xi}{}^{{\Sigma}{\Pi}}\partial_{{\Sigma}}  F_1 \, \; \partial_{{\Pi}} F_2=0 \; . \qquad
\end{eqnarray}

For $E_{7(7)}$ EFT, in which ${\Sigma}, {\Pi}, \Lambda , \Xi =1,...,56$,   these section  conditions can be presented in a simpler form
\begin{eqnarray}\label{sec=tdFdF2=0}
& t_{G}{}^{{\Sigma}{\Pi}}\partial_{{\Sigma}}  F_1 \, \; \partial_{{\Pi}} F_2=0 \; , \qquad \\
\label{sec=Sp2=0}
&  \Xi^{{\Sigma}{\Pi}}\partial_{{\Sigma}}  F_1 \, \; \partial_{{\Pi}} F_2=0 \; , \qquad
\end{eqnarray}
where $\Xi{}^{{\Pi}{\Lambda}}=- \Xi{}^{{\Lambda}{\Pi}}$ is the $Sp(56)$ symplectic `metric', $t_{G}{}^{{\Sigma}{\Pi}} =
\Xi{}^{{\Pi}{\Lambda}}t_{G\; {\Lambda}}{}^{{\Sigma}}$  and $t_{G\; {\Lambda}}{}^{{\Sigma}}$  are $E_{7(+7)}$ generators in {\bf 56} representation, $G=1,..., 133$.

To make the equations lighter, one usually writes the strong and the weak section conditions in the schematic form
\begin{eqnarray}\label{sec=Ydxd=0}
& Y_{\Lambda\Xi}{}^{{\Sigma}{\Pi}}\partial_{{\Sigma}} \otimes \partial_{{\Pi}} =0 \; , \qquad
\end{eqnarray}
and
\begin{eqnarray}\label{sec=Ydxd=0}
& Y_{\Lambda\Xi}{}^{{\Sigma}{\Pi}}\partial_{{\Sigma}} \, \partial_{{\Pi}} =0 \; . \qquad
\end{eqnarray}

It is natural to expect that the solutions of the section conditions  imply independence of all the fields on some number of internal coordinates. In the above schematic notation this can be expressed as
 \begin{eqnarray}\label{dy=Kdr}
 \partial_{{\Sigma}} (\ldots)= K_{\Sigma}{}^r \partial_r(\ldots)\; , \qquad \partial_r= \frac {\partial } {\partial \tilde{y}^r} \; , \qquad r=1,..., \tilde{n}_n\;
\end{eqnarray}
where $y^r$ are
$\tilde{n}_n$ ($<N_n$)  additional coordinates the fields are allowed to depend on.
A possible choice of this latter defines a {\it section} (i.e. a particular solution of the section conditions).
The freedom in choosing among the possible sections makes the construction $E_{n(n)}$--invariant.

For all the EFTs the $n$-parametric and $(n-1)$--parametric solutions of the section conditions were found and shown to describe
D=11 and D=10 type IIB supergravity \cite{Hohm:2013pua,Hohm:2013vpa,Hohm:2013uia,Hohm:2014fxa,Musaev:2014lna,Ciceri:2014wya,Abzalov:2015ega} (in the majority of the cases the bosonic limit of SUGRA was actually discussed). From the generic String/M-theoretic perspective,  one should not expected the possibility to have a solution with functions depending on more than 11 bosonic coordinates. Although for lower $n$, e.g. for lowest $n=2$ case in \cite{Berman:2015rcc}, this is manifest,  for higher $n$ this expectation had been just a reasonable  conjecture till  recent  \cite{Bandos:2015rvs}, where it has been proved for  the case of d=4 $E_{7(7)}$ EFT. Furthermore, in \cite{Bandos:2015rvs} it was shown that the set of 133 section conditions of this EFT, Eqs. (\ref{sec=tdFdF2=0}),
\begin{eqnarray}\label{sec=tdd=0}
t_{G}{}^{{\Sigma}{\Pi}} \partial_{{\Sigma}} \otimes \partial_{{\Pi}}=0 \; , \qquad
{\Sigma}, {\Pi}=1,...,56\; , \qquad G=1,...,133\; ,
\end{eqnarray}
is reducible in the sense that one can extract such a set of 63 conditions that their solution automatically solves also the remaining relations.

To be more specific in this latter statement, it was shown in  \cite{Bandos:2015rvs} that the solutions of the set of $63$ relations (\ref{sec=tdd=0}) involving the generators of $SU(8)$ subgroup of $E_{7(+7)}$,
 \begin{eqnarray}\label{secH=tdd=0}
t_{H}{}^{{\Sigma}{\Pi}} \partial_{{\Sigma}} \otimes \partial_{{\Pi}}=0 \; ,  \qquad
{\Sigma}, {\Pi}=1,...,56\; , \qquad H=1,...,63\; ,
\end{eqnarray}
automatically solve also the remaining $70$  conditions
which involve the  generators of the coset $\frac{E_{7(+7)}}{SU(8)}$ ($t_{K}{}^{{\Sigma}{\Pi}} \partial_{{\Sigma}} \otimes \partial_{{\Pi}}=0 $, $K=1,...,70$) as well as the strong section conditions (\ref{sec=Sp2=0}) involving the symplectic metric $\Xi^{\Sigma\Pi}=- \Xi^{\Pi\Sigma}$
\footnote{In terms of the derivatives  in {\bf 27} and  $\mathbf{\overline{ 27}}$  of SU(8),
$ \partial_{{\Sigma}}=  \left( \partial_{ij},  \bar{\partial}{}^{ij}\right)$,  the (formal) solution of the section conditions (\ref{secH=tdd=0})  can be obtained \cite{Bandos:2015rvs} by SU(8) transformations from   $\partial_{ij}=\Gamma^I_{ij}\partial_I= \bar{\partial}{}^{ij} $, with real $\partial_I= \bar{\partial}_I$ ($I=1,...,7$) and SO(7) Gamma matrices $\Gamma^I_{ij}$. It is easy to check that this solved  the conditions (\ref{sec=Sp2=0}) which, in its manifestly $SU(8)$ invariant form, reads $\partial_{ij}\otimes \bar{\partial}{}^{ij} - \bar{\partial}{}^{ij} \otimes \partial_{ij}=0$. To show that this solves also $70$ equation with coset generators, $\partial_{[ij} \otimes \partial_{kl]} - {1\over 4!}\epsilon_{ijkli'j'k'l'} \bar{\partial}{}^{i'j'}\otimes \bar{\partial}{}^{k'l'} =0$,  is a bit more involving
\cite{Bandos:2015rvs}.}.

To obtain the above results, it was very useful to analyze the classical mechanic counterpart of the section conditions which reads
\begin{eqnarray}\label{sec=tpp}
t_{E}{}^{{\Sigma}{\Pi}} \, p_{{\Sigma}}  \, p_{{\Pi}}=0\; ,  \qquad
\end{eqnarray}
where $ p_{{\Sigma}} $ and  $p_{{\Pi}}$ are classical momenta of a particle model. One notices that, if we perform a straightforward `quantization' of (\ref{sec=tpp}) by replacing the momentum by derivative, $
 p_{{\Sigma}} \mapsto -i \partial_{{\Sigma}}$,
consider (\ref{sec=tpp}) as a (first class) constraint and impose
its quantum version as a condition on the wave function, we clearly arrive at the weak version of (\ref{sec=tdd=0}) imposed on one function rather than on the pair of functions of EFT.
However, as it was discussed in \cite{Bandos:2015rvs}, there exists another 'first solve than quantize' way
which, starting from the classical section conditions (\ref{sec=tpp}) results in the (general solution of the) strong section conditions (\ref{sec=tdd=0}).
The key point is that such a general solution is expected to be of the form of (\ref{dy=Kdr}) and hence can be reproduced by quantization of the general solution of the classical section conditions of the form
\begin{eqnarray}\label{py=Kpr}
 p_{{\Sigma}} = K_{\Sigma}{}^r p_r\; , \qquad r=1,..., \tilde{n}_n\; .
\end{eqnarray}

The classical counterpart of the section conditions had been also studied in  \cite{Cederwall:2015jfa}  devoted to development of a twistor approach to  $E_{n(n)}$ EFTs with  $n\leq 6$.  In particular, in \cite{Cederwall:2015jfa}
it was discussed the classical section conditions of the $E_{4(4)}=SL(5)$  EFT which reads
\cite{Berman:2011cg,Blair:2013gqa}
\begin{eqnarray}\label{sec=tpp5}
p_{[\tilde{\mathfrak{a}}\tilde{\mathfrak{b}}}p_{ \tilde{\mathfrak{c}} \tilde{\mathfrak{d}}}=0\;  , \qquad \tilde{\mathfrak{a}},\tilde{\mathfrak{b}}, \tilde{\mathfrak{c}}, \tilde{\mathfrak{d}}=1,...,5 \; .  \qquad
\end{eqnarray}
Here $p_{\mathfrak{a}\mathfrak{b}}=-p_{\mathfrak{b}\mathfrak{a}}$ are momenta conjugate to the additional bosonic coordinates of the spacetime of the $E_{4(4)}=SL(5)$  EFT, $y^{\tilde{\mathfrak{a}}\tilde{\mathfrak{b}}}=-y^{\tilde{\mathfrak{b}}\tilde{\mathfrak{a}}}$ which belongs to {\bf 10} representation of $SL(5)$. The simple form of this SL(5) section conditions will be suggestive for our discussion below.

\subsection{Some differences between $E_{4(4)}$, $E_{7(7)}$ and $E_{8(8)}$ EFTs }

This is the place to illustrate the differences in the structure of section conditions of $E_{n(n)}$ EFTs with  different $n$.

First notice that, as it was shown in \cite{Blair:2013gqa} the solution of section conditions $\partial_{[\mathfrak{a}\mathfrak{b}}\otimes \partial_{\mathfrak{c}\mathfrak{d}]}=0$ corresponding to the embedding of D=11 and of  type IIB supergravity in the d=7 $E_{4(4)}$  EFT are independent in the sense that they are not connected by transformations of $E_{4(4)}=SL(5)$ group. The same applies to the classical counterparts of this strong section conditions given in Eq.
(\ref{sec=tpp5}). In contrast, as it can be deduced from the results of \cite{Bandos:2015rvs}, in the case of d=4 $E_{7(7)}$ EFT the situation is opposite: the solutions describing the embedding of D=11 and type IIB  supergravities into this EFT are related by transformations of the $SU(8)$ subgroup of $E_{7(7)}$.

Actually this distinction does not look unnatural after comparing  the number of bosonic  coordinates of $E_{n(n)}$  EFTs with that of the DFT. Indeed, a unification of the 11D and type  IIB solutions of a EFT implies also the unification of (low energy limits of the) type IIA and IIB superstring theories. This is reached in the frame of DFT which is defined in the space with doubled number of coordinates, $2D=20$.  From this perspective one can expect the independence of 11D and type IIB solution in $E_{n(n)}$ EFTs with $2 \leq n\leq 4$, where the number of additional and spacetime  coordinates is less that 20, and their unification in EFTs with $n\geq 5$. It will be interesting to check this hypothesis for $n=5,6$ and $n=8$ cases.

One more illustrative example is in difference between $E_{7(7)}$ and $E_{8(8)}$ EFTs.




The first is formulated in the space with 56 additional coordinates $y^\Sigma=({y}^{ij},\bar{y}_{ij})$ which can be considered \cite{Bandos:2015rvs} as a bosonic body of central charge superspace $\Sigma^{(60|32)}$ \cite{Howe:1980th}\footnote{The formulation of standard ${\cal N}=8$ $d=4$ supergravity in this superspace was described  in (Appendix B of)  \cite{Howe:1980th}  and in (Appendix C of) more recent  \cite{Howe:2015hpa}, where the central charge superspace formulation of other maximal d=11-n supergravities with $2\leq n\leq 8$ were  also considered.}.  The flat version of this superspace,  $\Sigma_0^{(60|32)}$, is the supergroup manifold associated with the most general central extension of the maximal $D=4$ ${\cal N}=8$ supersymmetry algebra
\begin{eqnarray}\label{SUSYal}
\{ Q_{\alpha}^i, Q_{\beta}^j\}= \epsilon_{\alpha\beta}Z^{ij}
\; , \qquad \{ Q_{\alpha}^i, \bar{Q}_{\dot{\beta}j}\}= \delta_j^i\sigma^a_{\alpha\dot{\beta}}P_a
\; , \qquad \{  \bar{Q}_{\dot{\alpha}i},  \bar{Q}_{\dot{\beta}j}\}= \epsilon_{\dot\alpha\dot\beta}\bar{Z}_{ij}
\; ,\qquad \\ \nonumber
\alpha,\beta =1,2\; , \qquad \dot{\alpha},\dot{\beta}=1,2\; , \qquad  a=0,1,2,3\; , \qquad i,j=1,...,8
\; . \qquad
\end{eqnarray}
This observation allowed us \cite{Bandos:2015rvs} to formulate a superparticle model in central charge superspace which generates
the classical counterpart of the independent section conditions (\ref{secH=tdd=0}) as a constraint.
The model is an improved version of the ${\cal N}=8$ superparticle described  by de Azc\'arraga and Lukiersi in   \cite{deAzcarraga:1984hf}. In the original model the invariance of the action under  $\kappa$--symmetry   can be reached only if we allow ourselves to impose  the  classical counterpart of independent section conditions,
(\ref{sec=tpp}) with $E=H$,
\begin{eqnarray}\label{sec=tpp=H}
t_{H}{}^{{\Sigma}{\Pi}} \, p_{{\Sigma}}  \, p_{{\Pi}} =0\; ,  \qquad H=1,..., 63\; ,
\end{eqnarray}
  `by hand', while in our improved version (which is not apparently equivalent to the original model) these
  appear as equations of motion.

A natural wish is to find a similar superparticle model generating (an independent part of) the section conditions for $E_{8(8)}$ EFT. But here we meet a problem already at the first stage. The number of central charges $Z^{pq}=-Z^{qp}$ of the central extention of maximal $d=3$ supersymmetry algebra
\begin{eqnarray}\label{susyM=3d}{}\{ Q^q_{\tilde{\alpha}}, Q^p_{\tilde{\beta}}\} = \gamma^{\tilde{a}}_{\tilde{\alpha}\tilde{\beta}} \delta^{pq} P_{\tilde{a}}+ i\epsilon_{\tilde{\alpha}\tilde{\beta}}Z^{pq}\; , \quad \tilde{\alpha}, \tilde{\beta} =1,2\; , \quad {\tilde{a}}=0,1,2\; , \quad p,q=1,..,16\end{eqnarray} is 120, while the number of the additional coordinates of $E_{8(8)}$ EFT is 248 \cite{Hohm:2013pua}. This is the dimension of the minimal irreducible representation of $E_{8(8)}$ which in \cite{Hohm:2013pua} was taken to be the adjoint representation.

Thus the relation  of additional coordinates of EFT with central charges of maximal central extension of  the maximal  $d$-dimensional supersymmetry algebra observed for  $E_{7(7)}$ EFT in \cite{Bandos:2015rvs} cannot be generalized straightforwardly to the $E_{8}$ case.

The idea of our study is  to insist nevertheless on the beautiful relation of  additional coordinates of EFT and of the maximal $d$ dimensional supersymmetry algebra. As we will see in a moment, this leads us to the conjecture on the existence of an underlying EFT (uEFT) 'living' in the  maximal {\it tensorial} central charge superspace. This can be defined at any $d\leq 11$, but its associated supersymmetry algebra always
has a hidden symmetry including $SO(1,10)$ so that it can be called M-algebra or M-theory superalgebra and
our uEFT can be called 11D EFT.

\subsection{$E_{8(8)}$ EFT and maximal supersymmetry}

If we insist on relation of additional coordinates of the $E_{n(n)}$ EFT with maximal $d=11-n$ supersymmetry algebra, in the case of $n=8$, $d=3$ we have to allow for contributions of some additional
coordinates carrying both the indices of the internal symmetry  SO(16) and of the d=3 Lorentz symmetry. Namely, we need in a coordinates conjugate to 128 of possible 405 additional vectorial 'central' charges $Y_a^{pq}=Y_a^{((pq))}$
(where double brackets imply symmetric traceless part:
$Y_a^{pq}=Y_a^{qp}$, $Y_a^{qq}=0$).

With the contribution of all  these generators the right hand side ({\it r.h.s.}) of the defining relation of the maximal $d=3$ supersymmetry algebra,
\begin{eqnarray}\label{susyA3d}
{}\{ Q^q_{\tilde{\alpha}}, Q^p_{\tilde{\beta}}\} = \gamma^{\tilde{a}}_{\tilde{\alpha}\tilde{\beta}}(\delta^{pq} P_{\tilde{a}} + Y_{\tilde{a}}^{((pq))})+ i\epsilon_{\tilde{\alpha}\tilde{\beta}}Z^{pq}\; , \qquad \\ \nonumber \tilde{\alpha}, \tilde{\beta} =1,2\; , \quad {\tilde{a}}=0,1,2\; , \quad p,q=1,..,16
\end{eqnarray}
becomes the generic $528$ component 32$\times 32$ matrix (528=3+405+120).

The mechanism of extraction of 128(=248-120)  additional coordinates of $E_{8(8)}$ EFT of  405(=528-3-120) additional coordinates conjugate to the  vectorial central charge of (\ref{susyA3d}) should be dynamical. A search for it is beyond the scope of this paper. For our discussion here the presence of even more
('beyond the  $E_{8(8)}$ EFT') additional coordinates is not problematic but rather suggestive.

Indeed at  this stage it is tempting to conjecture  the existence of an underlying exceptional field theory (uEFT), which includes as a sub-sectors  all the  $E_{n(n)}$ EFT  with $2\leq n\leq 8$ and lives in an enlarged superspace $\Sigma^{(528|32)}$ with  32 fermionic  and 528 bosonic coordinates. In terms of the above discussed $n=8$ case, these latter can be split on    $d=11-n=3$ spacetime, $120$ central charge and $405$ 'vector central charge' coordinates. But actually the similar splitting is possible for any $n$: the number of spacetime coordinates will be  $d=11-n$ while the set of additional $528-d$ coordinates will be split, in an $SO(1,d-1)$ invariant way, on the subsets of   scalar, vector and tensorial 'central' charge coordinates. The reason beyond this lays in a huge hidden automorphism symmetry of the superalgebra (\ref{susyA3d}) which we are going to discuss now.

\section{Maximal 11D tensorial central charge superspace $\Sigma^{(528|32)}$ and uEFT conjecture}

\subsection{ $\Sigma^{(528|32)}$ geometry and maximal supersymmetry}

The manifest $SO(1,2)\times SO(16)$ symmetry of  (\ref{susyA3d}) is related to the basis we have used to decompose the matrix of generators in {\it r.h.s.} of this relation. There exists also the  manifestly $SO(1,10)$ invariant form of the same relation,
\begin{eqnarray}\label{susyA11D}
{}\{ Q_{{\alpha}}, Q_{{\beta}}\} = i \Gamma^{{a}}_{{\alpha}{\beta}} P_{{a}} + \Gamma^{{a}{b}}_{{\alpha}{\beta}} Z_{{a}{b}} +i \Gamma^{{a}{b}{c}{d}{e}}_{{\alpha}{\beta}} Z_{{a}{b}{c}{d}{e}}
\; , \qquad
\\ \label{a,al,be=11D} \nonumber
a,b,c =0,...,9, 10 \; , \qquad \alpha,\beta,\gamma =1,...,32\; . \qquad
\end{eqnarray}
which explains the name of {\it M-algebra} or {\it M-theory superalgebra} \cite{Townsend:1995gp,Sorokin:1997ps}   often  used for this most general supersymmetry superalgebra\footnote{ Two comments are in time.  Firstly, the algebra  (\ref{susyA11D}) was described much before the M-theory epoch in  \cite{D'Auria:1982nx} and \cite{vanHolten:1982mx}. Secondly, in \cite{Sezgin:1996cj} the name 'M-algebra' was used for the superalgebra with additional fermionic generators. }.
The generators $Z_{{a}{b}}=Z_{[{a}{b}]}$ and $Z_{{a}{b}{c}{d}{e}}=
Z_{[{a}{b}{c}{d}{e}]}$ are called {\it tensorial central charges}.

Actually, the M-algebra possesses   $GL(32)$ automorphism symmetry which becomes manifest if we write it in the form
\begin{eqnarray}\label{susyASp}
{}\{ Q_{{\alpha}}, Q_{{\beta}}\} = i{\cal P}_{{\alpha}{\beta}}\; , \qquad {\alpha}, {\beta} =1,2,...,32\; , \qquad
\end{eqnarray}
collecting all the generators in the  {\it r.h.s.} in one symmetric $32\times 32$ matrix ${\cal P}_{{\alpha}{\beta}}$ ($528=\frac {32\times 33} 2$). Decomposing this on the basis of 11D gamma matrices and their products, \begin{eqnarray}\label{cP=11D} {\cal P}_{{\alpha}{\beta}}=  \Gamma^{{a}}_{{\alpha}{\beta}} P_{{a}} - i\Gamma^{{a}{b}}_{{\alpha}{\beta}} Z_{{a}{b}} + \Gamma^{{a}{b}{c}{d}{e}}_{{\alpha}{\beta}} Z_{{a}{b}{c}{d}{e}}\; , \qquad
\end{eqnarray} we arrive at  the form (\ref{susyA11D}) of the M-algebra, in which only the $SO(1,10)$ symmetry is manifest. The transformations from the
$GL(32)/SO(1,10)$ coset mixes the vector and antisymmetric tensor central charges among themselves.

If we complete the D=11 superspace by introduce the coordinates dual to every tensorial central charge generator, we arrive at superspace $\Sigma^{(528|32)}$ which is the supergroup manifold corresponding to the
maximal supersymmetry algebra  (\ref{susyASp}) or (\ref{susyA11D}). We denote coordinates of this superspace by
\begin{eqnarray}\label{ZcM=xyth}
{\cal Z}^{\frak M} &=& (X^{{\alpha}{\beta}}, {\theta}^{{\alpha}})
=
(x^{{a}}, y^{{a}{b}}, y^{{a}{b}{c}{d}{e}}, {\theta}^{{\alpha}})
\; , \\ \label{X=xyy}
& & X^{{\alpha}{\beta}}=X^{{\beta}{\alpha}}=
{1\over 32}\, x^{{a}}\, \tilde{\Gamma}_{{a}}^{{\alpha}{\beta}}- {i\over 64} \, y^{{a}{b}}\,\tilde{\Gamma}_{{a}{b}}^{{\alpha}{\beta}}  + {1\over 32\cdot 5!}\, y^{{a}{b}{c}{d}{e}}
\,\Gamma^{{a}{b}{c}{d}{e}}
_{{\alpha}{\beta}}\; . \qquad
\end{eqnarray}
The supersymmetric invariant Cartan forms of $\Sigma^{(528|32)}$ can be collected in a simple expressions
\begin{eqnarray}\label{Pialbe=}
\Pi^{{\alpha}{\beta}}&=& dX^{{\alpha}{\beta}} - i d{\theta}^{(\alpha}\; {\theta}^{\beta )}\; , \qquad \Pi^{{\alpha}}=d{\theta}^{{\alpha}}\; ,
\end{eqnarray}
which are covariant under $GL(32)$.
The $SO(1,10)$ invariant decomposition of the bosonic form reads 528=11+55+462, i.e. (see \cite{Bandos:2004ym} for properties of 11D gamma matrices in our notation)
\begin{eqnarray}\label{Pialbe=Pia+}
& & \Pi^{{\alpha}{\beta}}=
{1\over 32}\, \Pi^{{a}}\, \tilde{\Gamma}_{{a}}^{{\alpha}{\beta}}- {i\over 64} \, \Pi^{{a}{b}}\,\tilde{\Gamma}_{{a}{b}}^{{\alpha}{\beta}}  + {1\over 32\cdot 5!}\, \Pi^{{a}{b}{c}{d}{e}}
\,\Gamma^{{a}{b}{c}{d}{e}}
_{{\alpha}{\beta}}\; , \qquad
\end{eqnarray}
where
\begin{eqnarray}\label{Pialbe=Pia+}
 \Pi^a=dx^a - i d\theta^{\alpha}\Gamma^a_{{\alpha}{\beta}}\theta^{\beta}\; , \qquad \Pi^{ab}=dy^{ab} -  d\theta^{\alpha}\Gamma^{ab}_{{\alpha}{\beta}}\theta^{\beta}\; , \qquad \nonumber \\
 \Pi^{abcde}=dy^{abcde} -  id\theta^{\alpha}\Gamma^{abcde}_{{\alpha}{\beta}}\theta^{\beta}\; . \qquad
\end{eqnarray}

Our discussion above suggests to try to use the curved D=11 tensorial central charge superspace $\Sigma^{(528|32)}$ as an arena for constructing the {\it 11D EFT}, underlying the '$d=11-n$ dimensional' $E_{n(n)}$   EFTs with $n\leq 8$ (hence the name {\it uEFT} which we  also use for this 11D EFT).

The above described enlarged 11D superspace $\Sigma^{(528|32)}$ with additional  tensorial central charge coordinates, $y^{ab}$ and  $y^{abcde}$ in  (\ref{X=xyy}), was discussed in different contexts in \cite{Curtright:1987zc,Duff:1990hn,Gunaydin:1992zh,Bandos:1998vz,Bandos:1998wj,
Bandos:2001pu,Bandos:2003ng,Bandos:2004xw,Bandos:2004ym,Bandos:2005rr}. Of course, its $528$ bosonic coordinates can be considered as  finite subset of the infinite set of tensorial coordinates which were introduced  in \cite{West:2003fc} in the frame  of $E_{11}$ proposal \cite{West:2001as}--\cite{Tumanov:2016dxc} (see concluding section \ref{Conclusion} for more discussion on this). Notice also the relation of $\Sigma^{(528|32)}$ with hidden gauge symmetry \cite{D'Auria:1982nx,Bandos:2004xw,Bandos:2004ym} of 11D supergravity \footnote{In \cite{Duff:1990hn} the coordinates $y^{ab}$ were introduced to describe the duality transformations of supermembrane, and their possible relation with $E_{n(n)}$ duality symmetries was  discussed. In this respect it looks interesting that, as it was found in \cite{Bandos:2004xw,Bandos:2004ym}, the  hidden gauge symmetry of 11D supergravity can be associated with a one-parametric family of superalgebras the associated supergroup manifold of which  generically includes
$\Sigma^{(528|32)}$, but  one of the elements of this family is associated with
a smaller enlarged 11D superspace, containing  $y^{ab}$, but not   $y^{abcde}$
coordinate.}  \cite{Cremmer:1978km}, and that in this context the $GL(32)$ symmetry of $\Sigma^{(528|32)}$ is also broken down to its $O(1,10)$ subgroup.

The useful fact for our discussion below is that the $D=4,6$ and $10$ counterparts of this maximally enlarged  superspace (\ref{ZcM=xyth}), $\Sigma^{(m(m+1)/2|m)}$ with $m=4,8,16$,   provide the arenas for constructing  free massless conformal higher spin theories  in $D=4,6,10$ dimensional spacetimes \cite{Plyushchay:2003gv} \cite{Bandos:1999qf,Vasiliev:2001dc,Vasiliev:2001zy,Plyushchay:2003gv,Gelfond:2003vh,
Bandos:2004nn,Bandos:2005mb,Florakis:2014aaa,Gelfond:2015poa,Skvortsov:2016lbh,Goncharov:2016wxw}, and that these theories do possess $GL(m)$ and, moreover, the generalized superconformal $OSp(1|2m)$ symmetries.

The pioneering contribution in this 'tensorial superspace' or 'hyperspace' approach to conformal higher spin theories  was \cite{Fronsdal:1985pd} by Fronsdal,  where the  space parametrized by 4$\times$4 symmetric spin-tensorial coordinates (which can be decomposed on 4-vector and antisymmetric tensor ones)  was proposed as a generalization of spacetime appropriate for the description of
4D massless higher spin fields, and $Sp(8)$ was considered as a generalized conformal symmetry.
In \cite{Gunaydin:1992zh} Gunaydin proposed to introduce generalized spacetime coordinates by Jordan algebras. In particular, he treated $Sp(2m)$ as conformal group of Jordan algebra of  real  symmetric $m\times m$  matrices and $OSp(1|2m)$ as generalized  superconformal symmetry of the corresponding generalized superspace.
The first dynamical model formulated in $\Sigma^{(528|32)}$ superspace was the ''eleven dimensional  superstring'' by Curtright \cite{Curtright:1987zc} (see \cite{Bandos:2003ng} for even more exotic superstring model in $\Sigma^{(528|32)}$).

Coming back to our 11D uEFT proposal, the following comment is in time.
We appreciate that the relation  $d=11-n$ might suggest  $d=1$ or $d=0$ EFT to be the underlying one. However, such hypothetical  EFTs should have  infinite dimensional symmetry groups  $E_{10}$ \cite{Julia:1982gx,Gebert:1994mv,Damour:2000hv,Damour:2002cu,Henneaux:2008nr} and
$E_{11}$ \cite{West:2001as,West:2003fc,West:2007mh,West:2010rv,Henneaux:2010ys,West:2014eza,Tumanov:2015iea,Tumanov:2015yjd,Tumanov:2016abm,Tumanov:2016dxc},
which seems to imply the necessity to introduce an infinite number of additional coordinates.
In contrast a huge but finite number of unconventional  coordinates in our  11D EFT (528) provides us with the resource for additional coordinates for all the $E_{n(n)}$ EFTs with $2\leq n\leq 8$ (and also for
10D DFT)   although do not make any of these U-duality symmetries manifest. More discussion on
  possible relation/complimentarity  of our  11D EFT and $E_{11}$ proposal can be found in  Sec. \ref{Conclusion}. In the next section we present  the possible section conditions of the hypothetical
 11D EFT.

\subsection{Section conditions of the hypothetical 11D EFT}

In this section we propose the set of section conditions which can used
to reduce the number of spacetime coordinates in the  hypothetical 11D EFT.

\label{sec=sec-11D}
The proposed set of additional coordinates of the hypothetical uEFT, $y^{ab}$ and $y^{abcde}$, resembles the variables $y^{\frak{a}\frak{b}}=-y^{\frak{b}\frak{a}}$ of the $E_{4(4)}=SL(5)$ EFT, with the evident difference that in our case antisymmetric tensor coordinate carry Lorentz group indices, the same as the usual vector coordinate $x^a$, $a=0,...,9,10$.
Then  the simple form of the section conditions for  $E_{4(4)}=SL(5)$ EFT
\cite{Berman:2011cg,Blair:2013gqa}, Eq. (\ref{sec=tpp5}), suggests to try the following candidate  section conditions for the hypothetical 11D uEFT:
\begin{eqnarray}\label{11Dsec=dpdq}
&& \partial_{[a_1\ldots a_k}  \otimes  \partial_{b_1\ldots b_l]} + \partial_{[b_1\ldots b_l}\otimes  \partial_{a_1\ldots a_k]} =0 \; , \qquad k,l=1,2,5   \; , \qquad (k,l)\not= (1,1)\; . \qquad
\end{eqnarray}
The classical mechanic counterparts of these relations are
\begin{eqnarray}\label{11Dsec=p1}
&& p_{[a} p_{bc]}=0 \; , \qquad p_{[a} p_{bcdef]}=0   \; , \qquad \\
\label{11Dsec=p2}
&& p_{[ab} p_{cd]}=0 \; , \qquad p_{[ab} p_{c_1...c_5]}=0   \; .  \qquad
\end{eqnarray}
One might want to add $p_{[a}p_{b]}=0$ and $ p_{[b_1...b_5} p_{c_1...c_5]}=0 $, but these are satisfied identically at the classical level.

The trivial solution of these section conditions, $ p_{ab}= 0= p_{c_1...c_5}$, should reduce the uEFT to 11D supergravity. We expect also to have solutions which correspond to embedding of $E_{n(n)}$ ETS with $n\leq 8$.

Actually it is not difficult to find
the general solution of the first two equations, (\ref{11Dsec=p1}). It reads
\begin{eqnarray}\label{11Dsec=solp1}
 p_{ab}=p_{[a}q_{b]}\; , \qquad p_{abcde}=p_{[a}q_{bcde]}\; , \qquad \; \,  \qquad
\end{eqnarray}
with arbitrary $q_b$ and $q_{bcde}=q_{[bcde]}$.
This solves also the remaining part of the  classical section conditions, (\ref{11Dsec=p2}).

The easiest way to impose this solution of the section conditions on a function on the bosonic  bosonic body $\Sigma^{(528|0)}$ of $\Sigma^{(528|32)}$ (i.e. to quantize the classical section conditions using 'first solve then quantize' method), passes through the  Fourier transform  with respect to $x^a$.The quantum version of (\ref{11Dsec=solp1}) imposed on the (wave)function $\Phi (p_c, y^{cd}, y^{cdefg})\equiv \Phi (p, y^{[2]}, y^{[5]}) $,
 \begin{eqnarray}\label{11Dsec=sold1}
\partial_{ab} \Phi (p, y^{[2]}, y^{[5]})&=&-ip_{[a}q_{b]} \Phi (p, y^{[2]}, y^{[5]}) , \qquad  \nonumber \\ { } \partial_{abcde}\Phi (p, y^{[2]}, y^{[5]})&=& -i p_{[a}q_{bcde]}\Phi (p, y^{[2]}, y^{[5]}), \qquad
\end{eqnarray}
is solved by
\begin{eqnarray}\label{11Dsec=sol=ex}
\Phi (p, y^{[2]}, y^{[5]})= \exp \{ -i y^{bc} p_{[b}q_{c]} - i  y^{bcdef} p_{[b}q_{cdef]}\}\; \phi (p_a,
q_a, q_{a_1a_2a_3a_4}). \quad
\end{eqnarray}

One can appreciate that, as it is defined in the above equations, $ \phi (p_a,
q_a, q_{a_1a_2a_3a_4})$ dependence on  $q_a$ and $q_{abcd}$ should be such that the following redefinition of these do not change  $ \phi (p_a,
q_a, q_{a_1a_2a_3a_4})$ :
\begin{eqnarray}\label{q=q+p}
q_a \sim q_a+ \tilde{q} p_a \; , \qquad  q_{abcd} \sim  q_{abcd}  + \tilde{q}_{[bcd}  p_{a]}\; .
\end{eqnarray}
Taking into account that in this equations the second  'symmetry' is reducible,
\begin{eqnarray}\label{tq=tq+p}
 \tilde{q}_{abc} \sim  \tilde{q}_{abc}  +  \tilde{\tilde{q}}_{[ab} p_{c]}\; , \qquad  \tilde{\tilde{q}}_{ab} \sim   \tilde{\tilde{q}}_{ab}  +  \tilde{ \tilde{\tilde{q}}}_{[a} p_{b]}\; ,
\end{eqnarray}
one finds that effectively  $ \phi (p_a,
q_a, q_{a_1a_2a_3a_4})$ depends on $11-1+\large\{_{\;4}^{11} \large\}-\large\{_{\;3}^{11} \large\}+\large\{_{\;2}^{11} \large\}-\large\{_{\;1}^{11} \large\}=219$ additional momenta. On first glance this might look damaging for our hypothesis, as 219 is clearly less than 248, the dimension of minimal irreducible representation of $E_{8(8)}$.
However, let us recall that also in the case of  $E_{8(8)}$ EFT one expects the general solution of its  section condition to allow dependence of the functions on not more than $8$ coordinates, while a dependence on
other $241$ coordinates is 'unphysical' but needed to provide a freedom in choosing section and thus
the  $E_{8}$  invariance.

This suggests that the way from uEFT to $E_{8(8)}$ EFT  might pass through generic function
$ \phi (p_a,q_a, q_{a_1a_2a_3a_4})$, depending in an arbitrary ('unphysical') manner on $11+330=341 > 248$ variables
$q_a$ and $ q_{a_1a_2a_3a_4}$, and assume that, at the intermediate stage, an independence on some part of these appears due to some (dynamical or imposed) reduction mechanism. In this paper we will not try to find such a mechanism but rather assume its existence and exploit further the consequence of the idea of possible existence of the 11D uEFT.

Below we will describe a set of superparticle models proposed in \cite{Bandos:1998wj} and show that these produce quite generic solutions of the above section conditions as equations of motion. One of these models possess the maximal number $32$ of supersymmetries and $31$ local fermionic $\kappa$--symmetries so that it has a properties of BPS preon \cite{Bandos:2001pu}. It is nevertheless different from
the original  'preonic superparticle' of \cite{Bandos:1998vz}, and this difference  results in breaking of  the generalized superconformal symmetry $OSp(1|64)$ characteristic for the model of \cite{Bandos:1998vz}.

But before, let us discuss briefly the consistency of the proposed uEFT conditions (\ref{11Dsec=dpdq}) with the (solutions of the) section conditions of  $E_{n(n)}$ EFTs with  $n=5,6,7,8$.

\subsection{uEFT section conditions and  $E_{n(n)}$ EFTs with $n=5,6,7,8$.}

Our proposition for the EFT section conditions, Eqs. (\ref{11Dsec=dpdq}), are inspired by the form of the section conditions of the $E_{4(4)}=SL(5)$ EFT,
\begin{eqnarray}\label{sec=SL5}
& \partial_{[\tilde{\mathfrak{a}}\tilde{\mathfrak{b}}}\otimes \partial_{\tilde{\mathfrak{c}}\tilde{\mathfrak{d}}]}=0\; , \qquad \tilde{\mathfrak{a}},\tilde{\mathfrak{b}}, \tilde{\mathfrak{c}}, \tilde{\mathfrak{d}}=1,...,5
\; .  \qquad
\end{eqnarray}
The natural question to ask is whether they are consistent with these of other $E_{n(n)}$ EFTs, with $n=5,6,7,8$.

The section conditions of  $E_{5(5)}=SO(5,5)$ and $E_{6(6)}$ EFTs have the form \cite{Hohm:2013pua,Abzalov:2015ega}
\begin{eqnarray}\label{sec=E5}
\gamma_{I}^{{\Sigma}{\Pi}}\partial_{{\Sigma}} \otimes \partial_{{\Pi}} =0
\; ,  \qquad I=1,\ldots, 9, 10\; , \qquad {\Sigma}, {\Pi}= 1,\ldots , 16\; , \qquad \cite{Hohm:2013pua} \\  \label{sec=E6}
d^{{\Lambda}{\Sigma}{\Pi}}\partial_{{\Sigma}} \otimes \partial_{{\Pi}} =0\; , \qquad
{\Lambda}, {\Sigma}, {\Pi}= 1,\ldots , 27\; , \qquad \cite{Abzalov:2015ega}
\end{eqnarray}
and the doubts in consistency of our uEFT section conditions with these might arise from the observation that their number,
$10$ and $27$, are less than the numbers of internal components of Eqs.   (\ref{11Dsec=dpdq}) with  $n=5$ and $6$, \begin{eqnarray}\label{11Dsec=d1in}
&&  \partial_{[\mathfrak{a}} \otimes \partial_ {\mathfrak{b}\mathfrak{c} ]} +  \partial_{[\mathfrak{b}\mathfrak{c}} \otimes \partial_{\mathfrak{a}]}=0\; ,   \qquad  \partial_{[\mathfrak{a}\mathfrak{b}} \otimes \partial_{\mathfrak{c}\mathfrak{d}]} =0\; , \qquad \mathfrak{a}, \mathfrak{b}, \mathfrak{c}=1,\ldots, n\; , \quad n=5,6, \qquad  \\ \label{11Dsec=d2in} &&
\left[\begin{matrix}
\partial_{\mathfrak{a} } \otimes \tilde{\partial}{} + \tilde{\partial}{}\otimes  \partial_{\mathfrak{a}} =0\; , & \qquad
\tilde{\partial}{} := \frac {1} {5!} \epsilon^{\mathfrak{b}_1\ldots \mathfrak{b}_5 }\partial_{\mathfrak{b}_1\ldots \mathfrak{b}_5 } \; , & \qquad  n=5 \; , \cr \partial_{\mathfrak{a} } \otimes \tilde{\partial}{}^{\mathfrak{a}} + \tilde{\partial}{}^{\mathfrak{a}}\otimes  \partial_{\mathfrak{a}} =0\; , & \qquad
\tilde{\partial}{}^{\mathfrak{a}} := \frac {1} {5!} \epsilon^{\mathfrak{a}\mathfrak{b}_1\ldots \mathfrak{b}_5 }\partial_{\mathfrak{b}_1\ldots \mathfrak{b}_5 } \; , & \qquad  n=6 \; , \end{matrix}\right. \qquad
\end{eqnarray}
$15$ and $36$.

However, as we have already commented, Eqs. (\ref{11Dsec=dpdq}) are reducible, and the same applies to  the subset of their  internal components (\ref{11Dsec=d1in}),
(\ref{11Dsec=d2in}). The general solution of these contains the branch
\begin{eqnarray}\label{11Dsec=inSol}
& \partial_{\mathfrak{a}\mathfrak{b}}\; ... = \partial_{[\mathfrak{a}} \left( K_{\mathfrak{b}]}...\right) \; , \qquad  \begin{matrix} \tilde{\partial}{}\; ... =0\; , \qquad n=5, \cr \tilde{\partial}{}^{\mathfrak{a}}\; ... =0\; , \qquad n=6, \end{matrix}
\end{eqnarray}
which implies a possible non-trivial  dependence of the functions  on at least  $n=11-d$ coordinates \footnote{Other branches are characterized by  $\partial_{\mathfrak{a}}=0$ and allow possible dependence on some part of tensorial internal coordinates $y^{\mathfrak{a}_1\mathfrak{a}_2}$ and
$y^{\mathfrak{a}_1 \ldots \mathfrak{a}_5}$ ($y^{\mathfrak{a}_1 \ldots \mathfrak{a}_5}=\epsilon^{\mathfrak{a}_1 \ldots \mathfrak{a}_5\mathfrak{b}} \tilde{y}_{\mathfrak{b}}$ for $n=6$ and $y^{\mathfrak{a}_1 \ldots \mathfrak{a}_5}=\epsilon^{\mathfrak{a}_1 \ldots \mathfrak{a}_5} \tilde{y}$ for $n=5$).}.
To see a possibility of dependence on more coordinates, let us discuss the  above solution with  $K_{\mathfrak{b}}=\tilde{\tilde{{\partial}}}_{\mathfrak{b}}$ being a derivative with respect to some additional $n$-vector coordinates $\tilde{\tilde{y}}^{{\mathfrak{b}}}$. This implies that the functions of uEFT  obeying the internal part of uEFT  section conditions  may depend nontrivially, besides $x^\mu$ ($\mu=0,...,(d-1)$, $d=11-n$) and $x^{\mathfrak{a}}$,  also on these $\tilde{\tilde{y}}^{{\mathfrak{b}}}\; $ \footnote{The mixed sector of the uEFT section conditions  (\ref{11Dsec=dpdq}),  which contains equations which carry both $d=11-n$ spacetime and $n$ dimensional internal indices, does not put additional restrictions on the dependence of functions on purely internal coordinates. The simples way to check this passes through the  classical section conditions, which in their mixed and spacetime parts are solved by
 $p_{a\mathfrak{b}}= \frac 1 2 (p_aq_{\mathfrak{b}}- p_{\mathfrak{b} }q_a)$, where  $q_{\mathfrak{b}}$ is an arbitrary $n$-vector, the  'classical' counterpart of $\tilde{\tilde{{\partial}}}_{\mathfrak{b}}$, and $q_a$ is an additional $d$-vector,  $p_{ab}=p_{[a}q_{b]}$ {\it etc.}  }.  This shows that   (\ref{11Dsec=d1in}),
(\ref{11Dsec=d2in}) are not more, but rather less restrictive in comparison with the standard section conditions of $E_{5(5)}=SO(5,5)$ and $E_{6(6)}$ EFTs, (\ref{sec=E5}), (\ref{sec=E6}), the general solutions of which are expected to allow for dependence on not more than 5 and 6 ($n=11-d$) additional coordinates respectively.

This conclusion can be easily generalized also for the case of $E_{n(n)}$ EFTs with $n=7, 8$ in which, as we have already mentioned, the solution of section conditions is also expected to allow (and in $n=7$ case is shown \cite{Bandos:2015rvs} to  allow) the dependence of the functions of EFT on not more than 11 coordinates (i.e. on not more than $n=11-d$ internal coordinates $y^{\mathfrak{a}}$ and
$d$ spacetime coordinate $x^\mu$). It is not difficult to see that the uEFT section conditions (\ref{11Dsec=dpdq}) have not a stronger, but rather a weaker effect.

Indeed, the preferable branch  of the general solution of  (\ref{11Dsec=dpdq})  can be formally described by
\begin{eqnarray}\label{11Dsec=gsol}
&&  \partial_{{a}{b}} \,\ldots  = \partial_{[{a}} \left( K_{{b}]}...\right)  , \qquad   \partial_{{a}_1\ldots {a}_5 }\, \ldots =\partial_{[{a}_1}
 \left( K_{{a}_2 \ldots {a}_5]}\, \ldots \right),  \qquad
{a}, {b}=0,\ldots, 9,10\; , \qquad
\end{eqnarray}
which implies nontrivial dependence on at least $11$ coordinates $x^a$. Again, to see the possible dependence on more coordinates, we can consider $K_a =\tilde{\partial}_a= \frac {\partial} {\partial\tilde{y}^a}$, $K_{abcd} =\tilde{\partial}_{abcd}=\frac {\partial} {\partial\tilde{y}^{abcd}}$ which implies the dependence of the wave function, besides $x^a$, also  on additional vector and antisymmetric tensor coordinates, $\tilde{y}^a$ and  $\tilde{y}^{abcd}$. Actually, the classical counterpart of the above described branch of the solution of the uEFT section condition has been discussed in the previous sec. \ref{sec=sec-11D}.

\subsection{Dynamical model generating (solutions of) the section conditions}

\label{sec=spart}
Having a candidate set of section conditions, first questions to answer is whether the corresponding EFT subject to this conditions has  nontrivial solutions and, if yes, whether these are meaningful in the perspective of String/M-theory.
In this sec. \ref{sec=spart} we address a  classical counterpart of the first of these problems: we search for  supersymmetric particle models in $\Sigma^{(528|32)}$ generating solutions of  the classical section conditions (\ref{11Dsec=p1}), (\ref{11Dsec=p2}). The meaning of these models will be the subject of the next Sec. 4.

\subsubsection{Preonic superparticle and conformally invariant section conditions }

The most known superparticle model in maximal tensorial central charge superspace is the 'preonic superparticle' of \cite{Bandos:1998vz}. Its action
\begin{eqnarray}\label{S=lldX}
&& S= \int d\tau \lambda_{\alpha}\lambda_{\beta} \Pi_\tau^{\alpha\beta}\equiv
\int d\tau \lambda_{\alpha}\lambda_{\beta}  (\partial_\tau X^{\alpha\beta}- i
\partial_\tau\theta^{(\alpha} \theta^{\beta )})  \;  \qquad
\end{eqnarray}
contains, besides the bosonic and fermionic coordinate functions, $X^{\alpha\beta}(\tau)=X^{\beta\alpha}(\tau)$ and $\theta^{\alpha}(\tau)$, also independent bosonic spinor field $\lambda_{\alpha}(\tau)$, $\alpha =1,...,32$.

Actually the model can be defined with arbitrary number $m$ of values of the indices,
$\alpha,\beta =1,..., m$, and for each value of $m$ it possesses a rigid symmetry under $OSp(1|2m)$ supergroup as well as  local $\frac {m(m-1)}{2}$ parametric bosonic symmetry ($b$-symmetry) and local $(m-1)$ parametric fermionic $\kappa$--symmetry \cite{Bandos:1998vz}. For $m=4,8, 16$ cases
$\alpha,\beta$ can be treated as $Spin(1,D-1)$ indices (i.e. $SO(1,D-1)$ spinor indices) of $D=4, 6, 10$ dimensional spacetime and the quantization of the corresponding model results in an infinite tower of free conformal higher spin fields in these dimensions  \cite{Bandos:1999qf,Bandos:2005mb}. The role of the generalized superconformal group
is played in this approach by $OSp(1|2m)$ supergroup with the bosonic body $Sp(2m)$ playing the role of generalized conformal group \cite{Fronsdal:1985pd,Bandos:1999qf,Vasiliev:2001dc,Vasiliev:2001zy}

For our original case of $m=32$, the action possesses $31$ $\kappa$-symmetries and this implies that the ground state of the model preserves all but one 11D spacetime supersymmetry i.e. possess the property of BPS preon of M-theory in the terminology of \cite{Bandos:2001pu} (see \cite{Bandos:2005rr} for a review). This is the reason to apply ({\it a posteriori})  the name 'preonic superparticle' to the model of \cite{Bandos:1998vz}. However, the quantization of the  $m=32$ ($D=11$) preonic superparticle results in a quantum state spectrum including state vectors with an indefinite
mass. The physical interpretation of such quantum states is obscure.

For the generic value of $m$ the ground state of the model (\ref{S=lldX}) preserves $(m-1)$ of $m$ supersymmetries of $\Sigma^{(\frac{m(m+1)}{2}|m)}$ superspace which allows us to apply the name `preonic  superparticle' also to the cases of $m=2,4,8,16$ when the treatment of $\alpha, \beta$ as spinor indices is possible.

The canonical momentum conjugate to the bosonic coordinate function of the preonic superparticle, $p_{\alpha\beta}:= \frac {\partial L} {\partial \partial_\tau X^{\alpha\beta}}$, is expressed through the bilinear of the bosonic spinors,
 \begin{eqnarray}\label{p=ll}
&& p_{\alpha\beta}= \lambda_{\alpha}\lambda_{\beta}\; .
\end{eqnarray}

This provides a general solution of a kind of $GL(n)$ invariant counterpart of the classical section conditions:
 \begin{eqnarray}\label{pap=0}
&& p_{\alpha[\beta}p_{\gamma] \delta}=0 \; .
\end{eqnarray}
The corresponding counterpart of weak section condition imposed on a (wave) function
\begin{eqnarray}\label{dadph=0}
&& \partial_{\alpha[\beta}\partial_{\gamma] \delta}\phi(X)=0 \;
\end{eqnarray}
gives the bosonic equation proposed by Vasiliev in \cite{Vasiliev:2001dc,Vasiliev:2001zy}.
For $n=4,8,16$ the solutions of this equation describe the tower of free massless bosonic  conformal higher spin fields in $D=4,6,10$ (see \cite{Bandos:2005mb} for D=6,10 cases).

However, the fact that  for  $m=32$ the meaning of this equation and of its solutions is unclear
defends  us from temptation to propose its 'strong' generalization
\begin{eqnarray}\label{daxd=0}
&& \partial_{\alpha[\beta}\otimes \partial_{\gamma] \delta} + \partial_{\delta[\gamma }\otimes \partial_{\beta]\delta}=0 \;
\end{eqnarray}
as a candidate strong section condition for our hypothetical  uEFT \footnote{Notice that the solutions of Eq. (\ref{dadph=0}), the form of which can be found in which in sec. 4.2, also solve the 'strong' condition (\ref{daxd=0}). This is a good illustration of the statement (in sec. 2.1)  that 'first solve than quantize' approach provides a solution of strong section conditions. }.

Decomposing the symmetric $32\times 32$ matrix of the   generalized momentum (\ref{p=ll}) of the
preonic superparticle model  on the basis of 11D Dirac matrices (see (\ref{X=xyy})), we find the corresponding vector momentum and its tensor counterparts read
\begin{eqnarray}\label{pap2p5=ll}
&& p_a=\lambda \tilde{\Gamma}_a\lambda \; , \qquad p_{ab}= i\lambda \tilde{\Gamma}_{ab}\lambda , \qquad p_{abcde}= \lambda \tilde{\Gamma}_{abcde}\lambda \; .
\end{eqnarray}
For the generic bosonic spinor $\lambda_\alpha $ these do not obey the relation
(\ref{11Dsec=p1}) and (\ref{11Dsec=p2}) which we have proposed as candidate section conditions for
the hypothetical 11D uEFT.  Thus we have to search for a different $\Sigma^{(528|32)}$ superparticle model to generate (solutions of) these.

\subsubsection{Preonic superparticle with composite bosonic spinor }

Curiously enough, a simple modification of the preonic superparticle model makes it to obey the proposed classical section conditions of uEFT, (\ref{11Dsec=p1}) and (\ref{11Dsec=p2}). To this end it is sufficient to make the fundamental bosonic spinor $\lambda_\alpha$ composite,
\begin{eqnarray}\label{l=lqv-q}
 \lambda_\alpha = \lambda^{+}_qv_{\alpha}^{-q}\; . \qquad
\end{eqnarray}
Here $\lambda_q$ is a 16 component bosonic vector (spinor of $SO(9)$), $q=1,...,16$, and $v_{\alpha q}^{\;\; -}$ is a set of
 $16$ Majorana spinors of $SO(1,10)$ ($\alpha=1,...,32$) constrained by (see \cite{Bandos:2007mi,Bandos:2007wm} and below for more details and references)
\begin{eqnarray}\label{v-q-q=}
2v_\alpha^{-q} v_{\beta}^{-q} = u_a^={\Gamma}^a_{\alpha\beta}\; , \qquad
v_\alpha^{-q} \tilde{\Gamma}_a^{\alpha\beta} v_{\beta}^{-p}= \delta^{qp}u_a^=\; ,
\qquad
v_\alpha^{-q} C^{\alpha\beta} v_{\beta}^{-p}=0
\; . \qquad
\end{eqnarray}
These constraints include the 11D gamma matrices $ \Gamma^a{}_{\alpha}{}^{\beta}$ and
charge conjugation matrix $ C^{\alpha\beta} $; they are both imaginary in our mostly minus notation while
\begin{eqnarray}\label{tG:=}
 \Gamma^a_{\alpha\beta}= \Gamma^a{}_{\alpha}{}^{\gamma}C_{\gamma\beta}\; , \qquad \tilde{\Gamma}_a^{\alpha\beta} =  C^{\alpha\gamma}\Gamma^a{}_{\gamma}{}^{\beta}\end{eqnarray} are real and symmetric.

The sign indices of spinors, $^\pm$, and vectors, $^\#$ ($= ^{++}$) and $^=$ ($=^{--}$),  indicate the weight of different variables under $SO(1,1)$ transformations which play an important role when  clarifying the group theoretical meaning of the constrained variables $v_{\alpha q}^{\; -}$; we will discuss this in the next section.

It is important that the constraints (\ref{v-q-q=}) imply that the vector $u_a^=$ is light-like,
  \begin{eqnarray}\label{u=2=0}
  u^{=a}u_a^==0\; . \qquad
\end{eqnarray}
Furthermore, we can show that, as a result of these constraints, both the spacetime momentum $p_a$ and the momenta conjugate to the tensorial central charge coordinates, $p_{ab}$ and $p_{abcde}$, are proportional to $u_a^=$,
  \begin{eqnarray}\label{p=u=}
  && p_a = u_a^= \, \rho^\# \; , \qquad \qquad \\
  \label{p2=qu=}
  && p_{ab}=  u_{[a}^= q^\#_{b]}\; , \qquad
 p_{abcde}= u_{[a}^= q^\#_{bcde]}\; ,
\end{eqnarray}
where
 \begin{eqnarray}\label{rho++}
\rho^\# =  \lambda^+_q\lambda^+_q \; , \qquad
\end{eqnarray}
and  $q^\#_{b}$ and $ q^\#_{bcde}$ are also certain bilinears of $\lambda^+_q$ (we describe them below).
It is easy to see that (\ref{p=u=}), (\ref{p2=qu=}) solve the candidate uEFT section conditions
(\ref{11Dsec=p1}) and (\ref{11Dsec=p2}).

Thus  we have shown that a solution of the candidate  section conditions (\ref{11Dsec=p1}) and (\ref{11Dsec=p2}) is generated by the generalized superparticle model with  the action
\begin{eqnarray}\label{S=lvlvdX}
&& S= \int d\tau \lambda^+_q v_{\alpha}^{-q} \, \lambda^+_p  v^{-p}_{\beta} \, \Pi_\tau^{\alpha\beta}\equiv
\int d\tau \lambda^+_q\lambda^+_p  v_{\alpha}^{-q}v^{-p}_{\beta}  (\partial_\tau X^{\alpha\beta}- i
\partial_\tau\theta^{(\alpha} \theta^{\beta )})  \;  \qquad
\end{eqnarray}
where $X^{\alpha\beta}(\tau)$ and $\theta^{\alpha}(\tau)$ are bosonic and fermionic coordinate functions describing the embedding of the superparticle worldline into 11D tensorial central charge superspace
$\Sigma^{(528|32)}$, $\lambda^+_q(\tau) $ are 16-component bosonic vectors (which can be considered as spinors of $SO(9)$) and $ v_{\alpha}^{-q}(\tau)$ is a set of bosonic variables constrained by
(\ref{v-q-q=}).

\subsubsection{Spinor moving frame variables}

\label{smf=subsec}

We have seen that the constraints (\ref{v-q-q=}) are useful as due to them the momenta conjugate to the 11D spacetime coordinate and to the tensorial central charge coordinates obey the classical section conditions  (\ref{11Dsec=p1}) and (\ref{11Dsec=p2}). Furthermore, they also imply that the projection of the worldline to 11D spacetime is light-like as  (\ref{v-q-q=}) result in (\ref{p=u=}), (\ref{u=2=0}) and, hence,
\begin{eqnarray}\label{pp=0}
p^ap_a=0 \;  . \qquad
\end{eqnarray}
This implies  that the quantum states of our dynamical systems are massless from the perspective of 11D spacetime.

However, at first glance, the meaning of the constraints (\ref{v-q-q=})  might look obscure.
To clarify this, let us first notice that the above constraints have a trivial solution \begin{eqnarray}\label{v-aq=1aq}
v_\alpha^{-q}= \delta_{\alpha}{}^{ q}= \left(\begin{matrix}  0 &  0 \cr 0 &  I_{16\times 16}\end{matrix} \right)
\qquad
\end{eqnarray}
for which (with an appropriate representation of the 11D gamma matrices) $u_a^{=(0)}=\delta_a^0 - \delta_a^{10}$ and the composed   bosonic spinor   (\ref{l=lqv-q}) has  16 vanishing components,
\begin{eqnarray}\label{l=lq1-q}
 \lambda^0_\alpha = \lambda^{+}_q \delta_{\alpha}{}^{ q} = \left(\begin{matrix}  0 \; 0\;  ... \;  0 \; \lambda_{1}^{+} \; \lambda_{2}^{+} \;  ... \; \lambda^{+}_{16}\end{matrix} \right)
 \; . \qquad
\end{eqnarray}

The solution (\ref{v-aq=1aq}) breaks the manifest SO(1,10) Lorentz symmetry of the constraints (\ref{v-q-q=})  to its $[SO(1,1) \otimes SO(9)]\otimes {\bb K}_9$ subgroup (see below for definition of $ {\bb K}_9$). The general solution is given by a Lorentz rotated version of (\ref{l=lq1-q}) in which the parameters of the Lorentz rotations are considered as additional dynamical variables.

A Lorentz rotation of 11D spinors are described by real 32$\times$ 32 matrix taking values in
the double covering of the 11D Lorentz group, $Spin(1,10)$,
\begin{eqnarray}\label{harmV-inL11}
 V_{\alpha}^{(\beta)}= \left(\begin{matrix} v_{\alpha}^{+{q}} , & v_{\alpha}^{-{q}}
  \end{matrix}\right) \in Spin(1,10)\; , \qquad \alpha=1,...,32\; , \quad  q=1,...,16
 \; . \qquad
\end{eqnarray}
When the elements of this matrix is considered as fields, in our case as 1--dimensional fields
$V_{\alpha}^{(\beta)}(\tau)= \left(\begin{matrix} v_{\alpha}^{+{q}} (\tau), & v_{\alpha}^{-{q}}
 (\tau) \end{matrix}\right)$, (\ref{harmV-inL11}) can be called {\it spinor moving frame matrix}.
This matrix and its counterpart with sign inverted, $-V_{\alpha}^{(\beta)}(\tau)$, are in two--to--one
correspondence with the {\it moving frame matrix}, which is $SO(1,10)$ valued matrix $ U_a^{(b)}(\tau)$
\begin{eqnarray}\label{Uab=in}
 U_a^{(b)} = \left( {1\over 2}\left( u_a^{=}+u_a^{\#}
 \right), \; u_a^I \, , {1\over 2}\left( u_a^{\#}-u_a^{=}
 \right)\right)\; \in \; SO(1,D-1)\; ; \qquad
\end{eqnarray}
this
 describes the moving frame attached to the worldline.

The correspondence is given by the conditions of Lorentz invariance of the Gamma matrices (see (\ref{tG:=}))
\begin{eqnarray} \label{VGVT=UG}
V\Gamma^{(a)} V^T =   {\Gamma}^b U_b^{(a)}\; , \qquad \\ \label{VTGV=UG} V^T
\tilde{\Gamma}_b  V =  U_b^{(a)} \tilde{\Gamma}_{(a)}  \; , \qquad
\end{eqnarray}
and of the charge conjugation matrix
\begin{eqnarray} \label{VCV=C}
VCV^T=C \; . \qquad
\end{eqnarray}

The splitting of the Lorentz group valued matrix  $U_{a}^{(b)}$ in (\ref{Uab=in}) is invariant under
$SO(1,1)\otimes SO(9)$ subgroup of   $SO(1,10)$, and  the condition $U_{a}^{(b)}\in  SO(1,10)$ implies the following conditions on the vectors forming this matrix (see \cite{Sokatchev:1985tc,Sokatchev:1987nk})
\begin{eqnarray}\label{u--u--=0}
u_{ {a}}^{=} u^{ {a}\; =}=0\; , \quad    u_{ {a}}^{=} u^{ {a}\,I}=0\; , \qquad u_{
{a}}^{\; = } u^{ {a} \#}= 2\; , \qquad
 \\  \label{u++u++=0} u_{ {a}}^{\# } u^{ {a} \#
}=0 \; , \qquad
 u_{{a}}^{\;\#} u^{ {a} I}=0\; , \qquad  \\  \label{uiuj=-} u_{ {a}}^{ I}
 u^{{a}J}=-\delta^{IJ}.  \quad
\end{eqnarray}

With the suitable representation for 11D gamma matrices, the conditions of correspondence between moving frame and spinor moving frame variables, (\ref{VGVT=UG}), (\ref{VTGV=UG}) and (\ref{VCV=C})
(equivalent to  $V_{\alpha}^{(\beta)}\in  Spin(1,10)$ and actually defining $U_{a}^{(b)}\in  SO(1,10)$)
can be split into the following set of constraints for the spinor moving frame variables (this is to say for   the rectangular blocks of  $V_{\alpha}^{(\beta)}\in  Spin(1,10)$)
\begin{eqnarray}\label{v-v-=Gu--}
2v_\alpha^{-q} v_\beta^{-q} = \Gamma^{a}_{\alpha\beta} u_a^{=} \quad & (a) & ,
\quad
v^{-q}\tilde{\Gamma}_av^{-p} = u_a^{=} \delta^{qp} \quad (b) ,   \quad v_\alpha^{-q} C^{\alpha\beta} v_{\beta}^{-p}=0  \quad (c), \;  \nonumber \\  & &   \\ \label{v+v+=Gu++}
2v_\alpha^{+q} v_\beta^{+q} = \Gamma^{a}_{\alpha\beta} u_a^{\#} \quad & (a) & ,
\quad
v^{+q}\tilde{\Gamma}_av^{+p} = u_a^{\#} \delta^{qp} \quad (b) ,  \quad  v_\alpha^{+q} C^{\alpha\beta} v_{\beta}^{+p}=0 \quad (c), \;  \nonumber \\  & & \\ \label{v-v+=GuI}
2v_{(\alpha|}^{-q} \gamma^I_{qp} v_{|\beta)}^{+p} = \Gamma^{a}_{\alpha\beta}
u_a{}^{I} \quad & (a) &, \qquad   v^{-q}\tilde{\Gamma}_av^{+p} = u_a^{I}
\gamma^I_{qp} \; ,  \qquad I=1,...,9 \; , \qquad (b)\nonumber \\  & &  v_\alpha^{+q} C^{\alpha\beta} v_{\beta}^{-p}=i\delta_{qp}
\quad (c)\; . \qquad
\end{eqnarray}

Clearly, the relations (\ref{v-v-=Gu--})  coincide with
(\ref{v-q-q=}). Notice that just this set of relations, i.e. Eqs. (\ref{v-q-q=}),  are invariant under local $O(16)$ transformations of $v_\alpha{}^-_q $,
\begin{eqnarray}\label{vv=uG-all}
v_\alpha{}^{-q}  \mapsto v_\alpha{}^{-p} {\cal O}_{pq}\; , \qquad {\cal O}_{pp'}{\cal O}_{qp'}=\delta_{qp}\qquad \Leftrightarrow\qquad {\cal O}_{qp}\in O(16)\; .
\end{eqnarray}
This symmetry is broken down to $Spin(9)$ by the constraints (\ref{v-v+=GuI}a,b) which involve the
$d=9$ Dirac matrices
\begin{eqnarray}\label{gI=9d}
\gamma^I_{qp}=\gamma^I_{pq}\; , \qquad (\gamma^I\gamma^J+ \gamma^J\gamma^I)_{qp}= \delta^{IJ}\delta_{qp}\; , \qquad q,p=1,...,16\; , \qquad I=1,...,9\; . \qquad
\end{eqnarray}

The manifest gauge symmetry of the complete set of constraints  (\ref{v-v-=Gu--})--(\ref{v-v+=GuI}) is
$SO(1,1)\times Spin(9)$,
\begin{eqnarray}\label{SO9-SO=v}
& v_\alpha^{-q}  \mapsto v_\alpha^{-p} S_{pq}e^{-\beta} \; , & \qquad  v_\alpha^{+q}  \mapsto v_\alpha^{+p} S_{pq}e^{+\beta} \; , \qquad
\\ \label{SO9-SO=u}
& u_a^{=}  \mapsto u_a^{=}e^{-2\beta} \; ,  & \qquad u_a^{\#}  \mapsto u_a^{\#}e^{+2\beta} \; ,  \qquad  u_a^{I}  \mapsto u_a^{J} {\cal O}^{JI}\; ,
\end{eqnarray}
where
\begin{eqnarray}\label{SO9-SO} && SS^T=I_{16\times 16}\; , \qquad S_{pp'}\gamma^I_{p'q'} S_{qp'}=\gamma^J_{qp}{\cal O}^{JI}  \qquad \Rightarrow\qquad  {\cal O}^{IK}{\cal O}^{JK}=\delta^{IJ} ,\nonumber
\\ && \qquad \Leftrightarrow\qquad S_{qp}\in Spin(9) \; , \qquad {\cal O}^{IJ}\in SO(9) \; . \qquad
\end{eqnarray}
These $SO(1,1)\times Spin(9)$ transformations also leave invariant the splittings (\ref{Uab=in}) of moving frame matrix  and (\ref{harmV-inL11}) of the spinor moving frame matrix on rectangular blocks $v_\alpha^{\pm q}$.  However, if we consider a dynamical model involving only one of these two blocks, $v_\alpha^{- q}$ in the case of our model, the gauge symmetry is enhanced up to  $[SO(1,1)\otimes SO(9)]\subset \!\!\!\!\!\!\times {\bb K}_9$, where ${\bb K}_9$ transformations are defined by
\begin{eqnarray}\label{K9=v}
& v_\alpha^{-q}  \mapsto v_\alpha^{-q}  \; , & \qquad  v_\alpha^{+q}  \mapsto v_\alpha^{+q}+ v_\alpha^{-p}  \gamma^I_{pq}\, k^{\# I} \; , \qquad
\\ \label{K9=u}
& u_a^{=}  \mapsto u_a^{=}\; ,  & \qquad u_a^{\#}  \mapsto u_a^{\#} +  2 u_a^{I} k^{\# I} + u_a^{=}\, k^{\# I} k^{\# I}  \; , \qquad
 u_a^{I}  \mapsto u_a^{I}  + u_a^{=} k^{\# I} \; ,
\end{eqnarray}

Thus, in a theory which is invariant under $SO(1,1)\otimes SO(9)$ transformations (\ref{SO9-SO=v}) and does not contain $v_\alpha^{+q}$, the set of spinor variables  $v_\alpha^{-q}$ constrainted by
(\ref{v-v-=Gu--}) (equivalent to (\ref{v-q-q=})) can be identified with homogeneous coordinate of the coset
$SO(1,10)/[SO(1,1)\otimes SO(9)]\subset \!\!\!\!\!\!\times {\bb K}_9$ which is isomorphic to a nine-sphere
${\bb S}^9$ \cite{Galperin:1991gk,Delduc:1991ir,Galperin:1992pz}
\begin{eqnarray}\label{v-q=S9}
 \{ v_\alpha^{-q}\}   = \frac{SO(1,10)}{[SO(1,1)\otimes SO(9)]\subset \!\!\!\!\!\!\times {\bb K}_9 } =
{\bb S}^9\; .
\end{eqnarray}
In the model where these $v_\alpha^{-q}$ can be treated as spinor moving frame variable,
this ${\bb S}^9$ can be recognized as the celestial sphere of the 11D observer \cite{Galperin:1991gk,Delduc:1991ir,Galperin:1992pz}.

Using the above constraints and their consequences, such as the unity decomposition
\begin{eqnarray}\label{I=v-qv+q}
 \delta^{\alpha}_\beta = iC^{\alpha\gamma} (v_\gamma^{+q}v_\beta^{-q}-v_\beta^{+q}v_\gamma^{-q}) \qquad \Leftrightarrow \qquad i v_\alpha^{+q}v_\beta^{-q}-iv_\beta^{+q}v_\alpha^{-q}=C_{\alpha \beta}\; ,
\end{eqnarray}
one can check that
\begin{eqnarray}\label{v-qG2v-p=}
 && v^{-q}\tilde{\Gamma}_{ab}v^{-p}
:= v_\alpha^{-q}\tilde{\Gamma}_{ab}^{\alpha\beta}v_\beta^{-p}= - 2i \, u_{[a}^=u_{b]}^I\; \gamma^{I\,qp} , \qquad \\ \label{v-qG5v-p=}
&& v^{-q}\tilde{\Gamma}_{abcde}v^{-p}= -4\,  u_{[a}^=u_{b}^Iu_{c}^Ju_{d}^Ku_{e]}^L\;  \gamma^{IJKL\, qp}\; . \qquad
\end{eqnarray}

For the generalized superparticle model (\ref{S=lvlvdX})
the canonical momenta conjugate
to the tensorial coordinate functions $y^{ab}$ and $y^{abcde}$ have the form of (\ref{pap2p5=ll}) with composite bosonic spinor (\ref{l=lqv-q}), so that (\ref{v-qG2v-p=}) implies
\begin{eqnarray}\label{p2=uuIgI}
 && p_{ab}=  2\, u_{[a}^=u_{b]}^I\;  \lambda^+\gamma^{I}\lambda^+ , \qquad \\ \label{p5=uu4g4}
&& p_{abcde}= -4\,  u_{[a}^=u_{b}^Iu_{c}^Ju_{d}^Ku_{e]}^L\;  \lambda^+\gamma^{IJKL}\lambda^+\; , \qquad
\end{eqnarray}
where $\lambda^+\gamma^{I}\lambda^+:= \lambda_q^+\gamma^{I\,qp}\lambda_p^+$, {\it etc}.
This set of equations has the form of the general solution (\ref{p2=qu=}) of the classical section conditions (\ref{11Dsec=p1}), (\ref{11Dsec=p2}) with
\begin{eqnarray}\label{qa=}q_a = u_{a}^I\; \frac{\lambda^+\gamma^{I}\lambda^+}{(\lambda^+\lambda^+)}\; , \qquad q_{abcd} = -4u_{a}^Iu_{b}^Ju_{c}^Ku_{d}^L\;  \frac{\lambda^+\gamma^{IJKL}\lambda^+}{(\lambda^+\lambda^+)}\; . \qquad
\end{eqnarray}

Hence we have shown that the preonic superparticle with composite bosonic spinor (\ref{l=lqv-q}), described by the action  (\ref{S=lvlvdX}), generates a solution of the classical counterparts
(\ref{11Dsec=p1}), (\ref{11Dsec=p2}) of the proposed section conditions  (\ref{11Dsec=dpdq}) of the hypothetical uEFT. In this sense we can say that (\ref{S=lvlvdX}) is (one of the) {\it uEFT superparticle}(s).

\subsubsection{A family of superparticle 'solving' the classical section conditions}

The next natural question is: are there more uEFT superparticle models?
In this section we present the family of superparticle models in  $\Sigma^{(528|32)}$  superspace, first described in \cite{Bandos:1998wj}, and show that each of these generates  a constraint solving the classical section conditions (\ref{11Dsec=p1}), (\ref{11Dsec=p2}) of the hypothetical 11D EFT. The actions of these models can be collected in the universal expression
\begin{eqnarray}\label{S=rpq}
&& S= \int d\tau \rho^{\#}_{qp} \, v_{\alpha}^{-q}v_{\beta}^{-p}\Pi_\tau^{\alpha\beta}\equiv
\int d\tau \rho^{\#}_{qp} \, v_{\alpha}^{-q}v_{\beta}^{-p} (\partial_\tau X^{\alpha\beta}- i
\partial_\tau\theta^{(\alpha} \theta^{\beta )})  \; , \qquad
\end{eqnarray}
in which $\rho^{\#}_{qp}=\rho^{\#}_{qp}(\tau)$ is a symmetric $16\times 16$ bosonic matrix field,  $ X^{\alpha\beta}= X^{\alpha\beta}(\tau)$ and $\theta^{\alpha}=\theta^{\alpha}(\tau)$ are 528 bosonic and 32 fermionic coordinate functions, the same as in (\ref{S=lldX}) and (\ref{S=lvlvdX}), and
$v_{\alpha}^{-q}=v_{\alpha}^{-q}(\tau)$ are the spinor moving frame variables (\ref{v-q=S9}) discussed in the sec. \ref{smf=subsec}.

One can consider the action (\ref{S=rpq}) as describing a class of superparticle models  the properties of which depend essentially on the rank of symmetric matrix $\rho^{\#}_{qp}$. Alternatively one can speak about  dynamical system with several branches determined by this rank.
 Of these, let us especially notice  the following  particular cases preserving minimal and maximal amount of supersymmetry:
    \begin{itemize}
\item    The case of rank 16 matrix with unity eigenvalues,
 \begin{eqnarray}\label{rpq=rI}
  \rho^{\#}_{pq}= \rho^{\#}\delta_{pq}\; ,
\end{eqnarray}
describes the massless 11D superparticle (sametimes called M$0$-brane), see \cite{Bandos:2007mi,Bandos:2007wm}.
This model has 16 $\kappa$--symmetries and, correspondingly, its ground state preserves one half of 32 spacetime supersymmetries.
\item
The case of rank 1 matrix
\begin{eqnarray}\label{rpq=ll}
  \rho^{\#}_{pq}= \lambda^+_q\lambda^+_p\; ,
\end{eqnarray}
as discussed below, correspond to a preonic superparticle model (in terminology of \cite{Bandos:2001pu}). It  possesses $31$ $\kappa$--symmetries and, hence, its ground state preserves all but one supersymmetries.
 \end{itemize}

Generically, if we restrict the model by requiring all the eigenvalues of  matrix $\rho^{\#}_{pq}$ of the rank $r$ to be positive,  it always can be written in the form
    \begin{eqnarray}\label{rpq=lsls}
  \rho^{\#}_{pq}= \lambda^{+s}_q\lambda^{+s}_p\; ,\qquad s=1,..., r\; .
\end{eqnarray}

Thus, without loss of generality (in practical terms, i.e. if not considering a problematic models) one can describe the branch of the dynamical system  (\ref{S=rpq}) with ${\rm rank} (\rho^{\#}_{pq})=r$  by
\begin{eqnarray}\label{S=lplq}
&& S^{(r)}= \int d\tau \lambda^{+s}_q\lambda^{+s}_p \, v_{\alpha}^{-q}v_{\beta}^{-p}\Pi_\tau^{\alpha\beta}  \; , \qquad s=1,..., r\; . \qquad
\end{eqnarray}
In this family $S^{(1)}$ is the preonic action, corresponding to (\ref{rpq=ll}), while the standard massless superparticle action is $S^{(16)}$ with $\lambda^{+s}_q=\sqrt{\rho^{\#}}\delta^{s}_q$.

The action (\ref{S=lplq}) is invariant under the (32-$r$)-parametric local fermionic $\kappa$--symmetry
\begin{eqnarray}\label{kappa=16-r}
\delta_\kappa X^{\alpha\beta}=i \delta_\kappa \theta^{(\alpha} \,  \theta^{\beta)}
 \; , \qquad \delta_\kappa v_{\alpha}^{-q} =0\; , \qquad \delta_\kappa \lambda^{+s}_q=0
 \nonumber \\ \label{kappa=th16-r}
 \delta_\kappa \theta^{\alpha} = \kappa^{+q} v_{q}^{-\alpha} + \kappa^{-\tilde{s}}  w^{\tilde{s}}_q v_{q}^{+\alpha}
\; , \qquad
 \tilde{s}=1,..., (16-r) \; . \qquad
\end{eqnarray}
Notice that Eqs.  (\ref{kappa=th16-r}) describes the general solution of the equation
\begin{eqnarray}\label{kappa=thv-}
 \delta_\kappa \theta^{\alpha}   v_{\alpha}^{-q}  \lambda^{+s}_q =0  , \quad \left\{ \begin{matrix} q=1,...,16\; ,\cr s=1,...,r  \end{matrix}\right\}  \qquad  \Leftrightarrow \qquad   \delta_\kappa \theta^{\alpha}   v_{\alpha}^{-q}  \rho^{\#}_{qp} =0 ,   \quad {\rm rank}(\rho^{\#}_{qp})=r\; . \nonumber \\ {}
\end{eqnarray}
To write this solution we have introduced the set of Spin(1,10) spinors  $v_{q}^{\pm \alpha}=\pm i C^{\alpha\gamma} v_\gamma^{\pm q}$ which obey (see (\ref{I=v-qv+q}))
\begin{eqnarray}\label{v-v-=0}
v_{q}^{+ \alpha}v_{\alpha}^{-q}  = \delta_{pq}\; , \qquad v_{q}^{- \alpha}v_{\alpha}^{-q}  = 0\; , \qquad
\end{eqnarray}
and  a set of 16--vectors  $w^{\tilde{s}}_q$ orthogonal to  $\lambda^{+s}_q$ \begin{eqnarray}\label{wtsls=0}
 w^{\tilde{s}}_q  \, \lambda^{+s}_q =0 \; ,
  \; , \qquad s=1,..., r  \; , \qquad  \tilde{s}=1,..., (16-r) \; . \qquad
\end{eqnarray} In other words, that  are $(16-r)$ null-vectors of the rank $r$ matrix $\rho^{\#}_{qp}=  \lambda^{+s}_q \lambda^{+s}_p$, $\;  w^{\tilde{s}}_q \rho^{\#}_{qp}=0$.

\bigskip

Let us calculate the canonical momentum conjugate to the bosonic coordinates in (\ref{S=rpq}).
 In the spin-tensor notation we obtain \begin{eqnarray}\label{pab=rpqvv} p_{\alpha\beta}=  \rho^{\#}_{qp} \, v_{\alpha}^{-q}v_{\beta}^{-p}\; . \end{eqnarray}  Using the constraints (\ref{v-q-q=}) we can find that this implies that the spacetime momentum of the system is a light-like 11-vector  \begin{eqnarray}\label{pa=rqqua--} p_a=\frac {\rho_{qq}}{32} u_a^= \qquad \Rightarrow \qquad  p_ap^a=0\; . \end{eqnarray}
Hence from the 11D spacetime perspective, any of the models (\ref{S=lplq}) describes a  massless particle or a set of massless particles.

Furthermore, using (\ref{v-qG2v-p=}) it is not difficult to show that the momenta conjugate to the tensorial coordinates have the form $$p_{ab}= u^=_{[a}q^\#_{b]}\quad and \quad p_{abcde}= u^=_{[a}q^\#_{bcde]}$$ with
\begin{eqnarray}\label{r-qa=}q_a =  u_{a}^I\; \frac{\lambda^{+r}\gamma^{I}\lambda^{+r}}{(\lambda^{+s}\lambda^{+s})}\; , \qquad q_{abcd} = -4u_{a}^Iu_{b}^Ju_{c}^Ku_{d}^L\;  \frac{\lambda^{+r}\gamma^{IJKL}\lambda^{+r}}{(\lambda^{+s}\lambda^{+s})}\; . \qquad
\end{eqnarray}
Thus any model from the family described by a (nondegenerate) action  of the form (\ref{S=rpq}) or (\ref{S=lplq}) generate a solution of the classical section conditions (\ref{11Dsec=p1}), (\ref{11Dsec=p2}) of the hypothetical underlining uEFT.

\section{On uEFT superparticles and 11D higher spin theories}

In the previous Section
\ref{sec=spart} we have presented a family of superparticle models which produce as constraints  quite generic solutions of the section conditions proposed for the hypothetical underlying 11D EFT (uEFT) in Sec. \ref{sec=sec-11D}.
In this section we will argue  that, curiously enough, the quantization of these uEFT superparticles should result in the theory of free massless higher spin fields in 11 dimensional spacetime.

\subsection{
Free $D=4,6,10$  conformal higher spin theory description in $\Sigma^{(\frac {m(m+1)}{2}|m)}$ superspace
with $m=2(D-2)=4,8,16$ }

To ague in favor of the above conclusion, we begin with already mentioned relation
of the original preonic superparticle model (\ref{S=lldX}) with $m=4,8$ and $16$ ($\alpha,\beta =1,...,m$)   with  free conformal massless higher spin field theories in spacetime of dimensions  $D=\frac {m+4} 2 = 4, 6, 10 $ \cite{Bandos:1999qf,Bandos:2005mb}. Namely, the quantization of these  models of superparticle in
$\Sigma^{(\frac {m(m+1)}{2}|m)}$ superspace
with $m=2(D-2)=4,8,16$  results in the quantum state spectrum described by an infinite tower of all D=4,6 and 10  massless conformal higher spin fields.

This is related to the fact that  generalized superconformal symmetry
$OSp(1|2m)$ can be realized  on  towers of the bosonic and of the fermionic massless conformal fields
which can be packed into a scalar $\phi(X)$  and a 'spinor' (s-vector) field $f_\alpha(X)$  on the  tensorial space (hyperspace) $\Sigma^{(\frac {m(m+1)}{2}|0)}$ (see \cite{Fronsdal:1985pd} for $m=4$)  which obey the Vasiliev's equations $\partial_{\alpha[\beta}\partial_{\gamma] \delta}\phi(X)=0$ (\ref{dadph=0}) and $\partial_{\alpha[\beta}f_{\gamma]} (X)=0$ \cite{Vasiliev:2001dc,Vasiliev:2001zy}. These fields can be also collected in superfield defined on $\Sigma^{(\frac {m(m+1)}{2}|m)}$ superspace satisfying   $D_{[\alpha}D_{\beta ]} \Phi(X,\theta)=0$ \cite{Bandos:2004nn}.

On the other hand, all the tower of the solutions of all the free conformal higher spin equations for bosonic fields in D=4,6 and 10 can be described by a scalar function $\tilde{\phi} (\lambda)$ of one unconstrained real bosonic spinor $\lambda_\alpha$, $\alpha =1,..., m$, with $m=2(D-2)$, subject to the restriction to be even with respect to $\lambda_\alpha \rightarrow - \lambda_\alpha$, and also by specifying the class of functions $\tilde{\phi} (\lambda)$  belongs to \cite{Bandos:1999qf,Bandos:2005mb}. With a suitable choice of this latter the solution of the bosonic Vasiliev equation (\ref{dadph=0}) is given by
 \begin{equation}\label{phi=int}
 {} \phi (X)= \int d\lambda \, \tilde{\phi} (X, \lambda) =\int d\lambda \, \tilde{\phi} (\lambda)\, e^{i\lambda_\alpha\lambda_\beta X^{\alpha\beta}}   {}
\; . \end{equation}

\subsection{m=4,8,10 counterparts of the preonic superparticle and conformal higher spin fields in
 D=4,6,10}

This $\tilde{\phi} (\lambda)$, and also  its fermionic counterpart, can be obtained by quantization \cite{Bandos:1999qf,Bandos:2005mb} of the $m=4,8,10$ verisons of the  superparticle model (\ref{S=lldX}) in terms of  components of orthosymplectic twistor
$(\lambda_\alpha , \mu^\alpha , \eta )$ related to the $\Sigma^{(\frac {n(n+1)}{2}|n)}$ coordinates by
 \begin{equation}\label{mu=Xl+}   \mu^\alpha =
 X^{\alpha\beta} \lambda_\beta - \frac i2 \theta^\alpha\theta ^\beta \lambda_\beta\; , \qquad \eta = \theta^\alpha \lambda_\alpha\; . \end{equation}
The  fundamental representation of $OSp(1|2m)$ acts on orthosymplectic supertwistors by left multiplication,  and the above incidence relations (\ref{mu=Xl+}) explain the possibility to realize $OSp(1|2m)$ as superconformal symmetry of $\Sigma^{(\frac {m(m+1)}{2}|m)}$.

For simplicity, we restrict our discussion here by quantization of purely bosonic limit, $\theta=0$, of  $m=2(D-2)=4,8,16$ superparticle (\ref{S=lldX}). Using the Leibniz rule the bosonic action in (\ref{S=lldX}),
\begin{eqnarray}\label{S0=lldX}
S_0= \int d\tau \lambda_\alpha\lambda_\beta \partial_\tau X^{\alpha\beta}\; ,
\end{eqnarray}
can be written in the form
\begin{eqnarray}\label{S0=ldmu} S= \int d\tau \left( \lambda_\alpha \partial_\tau \mu^\alpha - \partial_\tau \lambda_\alpha \; \mu^\alpha  \right)
\end{eqnarray}
with   $\mu^\alpha = X^{\alpha\beta} \lambda_\beta$ (\ref{mu=Xl+}). This new variable $\mu^\alpha (\tau)$
carries all the physical  degrees of freedom in $ X^{\alpha\beta}(\tau) = X^{\beta\alpha}(\tau) $ (the remaining
$\frac {m(m-1)} 2$ components  can be gauged away) and
can be considered as a momentum conjugate to $\lambda_\alpha $ (or {\it vise versa}: coordinate conjugate to momentum $\lambda_\alpha$) and  the action (\ref{S0=ldmu}) can be considered as a Hamiltonian action  with Hamiltonian equal to zero.

Then the  quantization of the model (\ref{S0=ldmu}) is trivial; its state vector can be represented by  an arbitrary function of $\lambda_\alpha$, $\tilde{\phi}(\lambda)$. The spacetime treatment of this quantum state spectrum  uses the relation (\ref{p=ll}),
 \begin{eqnarray}\label{p-ll=0}
&& p_{\alpha\beta}- \lambda_{\alpha}\lambda_{\beta}=0\; ,
\end{eqnarray}
which can be obtained as a primary constraint when constructing Hamiltonian approach to our dynamical system on the basis of the original action  (\ref{S0=lldX}).

An alternative quantization of (\ref{S0=lldX}) with $m=2(D-2)=4,8,16$, which passes through the stage of development of such a Hamiltonian approach and conversion of the second class constraints \cite{Bandos:1999qf},
results in a wavefunction dependent on both $X^{\alpha\beta}$ and $\lambda_\gamma$ and obeying the quantum counterpart of the constraint (\ref{p-ll=0}), the so--called preonic equation
 \begin{eqnarray}\label{dPhi=llPhi=0}
&& (\partial_{\alpha\beta}+ i  \lambda_{\alpha}\lambda_{\beta})\varphi(X,\lambda) =0\; .
\end{eqnarray}
The solution of this equation  is given by $\tilde{\phi} (X, \lambda) = \tilde{\phi} (\lambda)\, e^{i\lambda_\alpha\lambda_\beta X^{\alpha\beta}}$ and its integration with a suitable measure  $d^n\lambda$   give the wavefunction in the generalized coordinate, $X^{\alpha\beta}$ representation (\ref{phi=int}).

On the other hand, the wavefunction in the momentum  representation, $\phi (p_{\alpha\beta})$,  is localized on the solutions of (\ref{p-ll=0}),  which implies, in particular, that the standard $D$-vector momentum extracted from  (\ref{p-ll=0}) is
 \begin{eqnarray}\label{p=lGl}
&& p_a= \lambda\tilde{\Gamma}_a\lambda \equiv  \lambda_{\alpha}\tilde{\Gamma}_a^{\alpha\beta} \lambda_{\beta}\; .
\end{eqnarray}
For $D=4,6,10$ (and also for $D=3$) this momentum is light-like $p_ap^a=0$ and hence the quantum states of the model are massless.

Actually,  the quantum state spectrum of D=4 model consists of an infinite tower of the massless  fields of all possible helicities. In the case of D=6 and D=10 model, where, in contrast to $D=4$,  not all the free massless fields are conformal, we obtain a tower of massless conformal higher spin fields  (see e.g. \cite{Bandos:2005mb} for their description). The fields in the tower are 'enumerated' by a set of integer numbers which can be considered as momenta  conjugate to the coordinates of ${\bb S}^{(D-3)}$ (${\bb S}^{1}$, ${\bb S}^{3}$ and ${\bb S}^{7}$) spheres realized as Hopf fibrations $${\bb S}^{(n-1)}/{\bb S}^{D-2}= {\bb S}^{2D-5}/{\bb S}^{D-2} =\left({\bb S}^{3}/{\bb S}^{2}, {\bb S}^{7}/{\bb S}^{4}, {\bb S}^{15}/{\bb S}^{8}\right)\; . $$

Let us describe how these appear. The space of light-like momenta in D-dimensions is \begin{eqnarray}\label{p-space} \{p_a| p^2=0\} =  {\bb R}_+\otimes {\bb S}^{(D-2)}\;  \end{eqnarray} The space of nonvanishing $n$-component bosonic spinors our wavefunction $\tilde{\phi}(\lambda)$ depends on  is
\begin{eqnarray}\label{l-space} \{ \lambda_\alpha \} ={\bb R}^n - \{ 0\} =  {\bb R}_+\otimes {\bb S}^{(n-1)} \; ,   \end{eqnarray}
and the scale of momenta ($ {\bb R}_+$ in (\ref{p-space})) is given by the square of the scale of the bosonic spinor ($ {\bb R}_+$ in (\ref{l-space})). Thus, besides the light-like momenta, the wavefunction depend on $(D-3)$  coordinates of  the fibrations
\begin{eqnarray}\label{Hopf-f}
\{ \lambda_\alpha \} / \{p_a| p^2=0\} = {\bb S}^{(n-1)} /  {\bb S}^{(D-2)}= {\bb S}^{2D-5} /  {\bb S}^{(D-2)}\; , \qquad D=4,6,10 \; ,   \end{eqnarray}  which are isomorphic to ${\bb S}^{(D-3)}$ spheres.
These spaces are compact and a momentum conjugate to a compact coordinate is quantized.

Hence passing to the momentum representation on this compact directions,  we will arrive at the wave function  depending on D dimensional light-like momentum
($D=4,6,10$) and characterized by $(D-3)$ integer numbers. In D=4  one integer number obtained in such a way is the doubled  helicity of a massless fields.
The description of D=10 and D=6 conformal higher spin fields  can be found in \cite{Bandos:2005mb} and refs.  therein.

\subsection{On superparticle models for massless non-conformal higher spin theories in
D=6,10 and D=11}

Notice that, although the interest  in a tensorial (super)space or hyper(super)space description of higher spin fields persists already more than 15 years
\cite{Bandos:1999qf,Vasiliev:2001dc,Vasiliev:2001zy,Plyushchay:2003gv,Gelfond:2003vh,Bandos:2004nn, Bandos:2005mb,Florakis:2014aaa,Gelfond:2015poa,Skvortsov:2016lbh}, the research is   mainly concentrated on  D=4 case. The reason beyond this, besides that the $m=8,16$ cases are more complicated, is that, in contradistinction to D=4, in  D=6 and D=10 dimensional cases not all the massless fields are conformal, and  these  which are look quite exotic \cite{Bandos:2005mb}. In particular neither the linearized equations for graviton, nor  the  Maxwell  equations for D-vector potential are conformally invariant in D=6 and D= 10 dimensions.

In the 11D case the straightforward generalization of the above derivation of free higher spin theories  fails. Namely, symplectic twistor quantization is universal and results in a wavefunction $\tilde{\phi}(\lambda)$ depending on $m=32$ component bosonic spinor in an arbitrary manner, but what fails is its spacetime interpretation: in contrast to  D=4,6,10 cases, in D=11 the momentum  constructed from spinor bilinear (\ref{p=ll}) is not light-like. Actually with this 11-momentum
  the mass of the quantum states remains indefinite which hamper the  spacetime interpretation of D=11 (super)particle model (\ref{S0=lldX}) ((\ref{S=lldX})).

In the above perspective there exist the quests for  superparticle models providing the classical mechanic description of D=11 higher spin field theory and of D=6, 10 dimensional massless {\it non-conformal} higher spin theories. The  $m=32$ and $m=8, 16$ versions of the above described superparticle models (\ref{S=rpq}) and (\ref{S=lplq}) are good candidates for these roles.

This conjecture is suggested by a series of observations the first of which is  that, as we have described above, all these models produce the constraints $p_a\propto u_a^=$ which implies $p_ap^a=0$. As a result, their quantum state spectrum is formed by massless states. Then, the  analogy with the above discussed $n=2(D-2)=4,8,16$ version of the preonic superparticle model suggests that this quantum state spectrum provides us with a theory of free  higher spin fields.

This conjecture looks especially natural  in the case of preonic-type model (\ref{S=lvlvdX}) with composite spinor field (\ref{l=lqv-q}). As far as the other models (\ref{S=rpq}), (\ref{S=lplq}), preserving from one half to all but two supersymmetries are concerned, this  might be considered as a counterpart of the D=4 $OSp(4|2)$  invariant models in \cite{Bandos:1999qf} which preserve 2 of 4 supersymmetries and also describes supermultiplet of free massless higher spin fields by its quantum state spectrum.

Furthermore, as  there are no traces of conformal invariance in the models  (\ref{S=rpq}), (\ref{S=lplq}) with $m=8, 16, 32$, their quantum state spectrum should not be conformal.   Indeed,
it is easy to see that even in the preonic-type  model with composite bosonic spinor (\ref{S=lvlvdX}), which is included as $r=1$ representative in the set of models (\ref{S=lplq}), the presence of spinor moving frame variables $v_{\alpha q}^{\; -}$ in (\ref{l=lqv-q}) breaks the $Sp(m)$ invariance down to D-dimensional Lorentz group.

Thus we have argued that the quantization of the models (\ref{S=lplq}), (\ref{S=rpq}) with
$m=8,16$ and $32$
should result in a theory of free non-conformal higher spin fields in $D=6,10$ and $D=11$ dimensional spacetime.

The check of this conjecture by explicit quantization of these models and by the analysis of their quantum state spectrum will be the subject of a forthcoming paper. In the next section we present a discussion on some representation of the M-theory superalgebra and on the embedding of 11D supergravity in these representations,  which actually suggests  how the quantum state spectrum of some of these models might look like.

To conclude this section, we just notice that the idea on that the 11D higher spin fields are necessary ingredients of (the hypothetical) underlying 11D exceptional field theory, uEFT,  is in consonance
with discussions on their necessity in the context of (also hypothetical) $E_{10}$ and  $E_{11}$ theories   \cite{West:2007mh,Henneaux-Talk}.

\section{Embedding 11D supergravity into representations of the M-theory superalgebra, the supersymmetry superalgebra of uEFT}

One more argument in favor of relevance of 11D tensorial central charge superspace (\ref{S528-32}) as a basis for hypothetical underlying uEFT can be gained by discussing unitary representations of the M-theory superalgebra (\ref{susyA11D}), the supersymmetry superalgebra of $\Sigma^{(528|32)}$,  and by  showing that 11D supergravity multiplet can be included in (some of) these representations. A more complicated counterpart of such a study can be found in
 \cite{Gunaydin:1998bt} where Gunaydin showed that $osp(1|32)$ admits unitary representations which contain 11D SUGRA  when the contraction and reduction to the super-Poincar\'e is taken.

Actually, as far as M-algebra (\ref{susyA11D}) can be also obtained by contraction of  $osp(1|32)$, the affirmative answer on the question of whether
11D SUGRA can be embedded in some of its highest weight unitary representations  is guaranteed by the results of  \cite{Gunaydin:1998bt}. However, we find suggestive to construct such an embedding explicitly, in particular because it gives us a hint about how the results of the quantization of some of the superparticle models described in section 3 might look like.

The construction of highest weight representations of M-algebra (\ref{susyA11D}) is simpler than that of
semisimple superalgebra $osp(1|32)$:  the algebra of its  bosonic generators, $P_a$, $Z_{ab}$ and $Z_{abcde}$, which can be collected in
${\cal P}_{\alpha\beta}$ (\ref{susyASp}), is Abelian, $[{\cal P}_{\alpha\beta}, {\cal P}_{\gamma\delta}]=0$, and, hence, they can have the basis of common eigenvectors or eigenstates.
 We denote such eigenstates by $|{\cal A}, p_{\alpha\beta}>$, where  $p_{\alpha\beta}$ are  eigenvalues of ${\cal P}_{\alpha\beta}$,
\begin{eqnarray}\label{cPS=pS}
{\cal P}_{\alpha\beta} |{\cal A}, p_{\alpha\beta}>=  p_{\alpha\beta} |{\cal A}, p_{\alpha\beta}>\; ,   \end{eqnarray}
and ${\cal A}$ denotes possible indices or additional variables the state  depends on.

\subsection{A particular class of eigenstates of the generalized momentum}

Let us discuss a particular class of such states, $|{\cal A}, v_{\alpha q}^{\;-}, \lambda_q^{+s} >$, for which the eigenvalue  matrices $p_{\alpha\beta}$ have rank $r\leq 16$ and can be presented in the form ({\it cf.}
 (\ref{pab=rpqvv}) with (\ref{rpq=lsls}))
 \begin{eqnarray}\label{pab=rpqvvA} p_{\alpha\beta}=  \lambda_q^{+s} \lambda_p^{+s} \, v_{\alpha}^{-q}v_{\beta}^{-p}\; , \end{eqnarray}
 where $v_{\alpha}^{-q}$ form a rectangular (32$\times$ 16) block of a spin group valued matrix (see (\ref{harmV-inL11})), and hence  obeys
 (\ref{v-q-q=}), and $\lambda_q^{+s}$ is 16$\times r$ matrix of maximal rank;
 \begin{eqnarray}\label{cPS=vvllS}
{\cal P}_{\alpha\beta} \; |{\cal A}, v_{\alpha q}^{\;-}, \lambda_q^{+s} >=  \lambda_q^{+s} \lambda_p^{+s} \, v_{\alpha}^{-q}v_{\beta}^{-p} \; |{\cal A}, v_{\alpha q}^{\;-}, \lambda_q^{+s} >\; .   \end{eqnarray}

 Let us observe that (\ref{pab=rpqvvA}) and (\ref{v-q-q=}) imply that the eigenvalue of the 11--momentum operator
 $P_a$, $p_a\propto \Gamma_a^{\alpha\beta} p_{\alpha\beta} \propto \lambda_q^{+s} \lambda_q^{+s} u_a^=$ is light-like, $p_ap^a=0$,
   \begin{eqnarray}\label{PaS=vvllS}
{P}_{a} \; |{\cal A}, v_{\alpha q}^{\;-}, \lambda_q^{+s} >=  \lambda_q^{+s'} \lambda_q^{+s'} \, u_a^=  \; |{\cal A}, v_{\alpha q}^{\;-}, \lambda_q^+ >\; ,  \qquad u_a^{=}u^{a=}=0\; .  \end{eqnarray}
Furthermore, as (\ref{pab=rpqvvA}) is invariant under $SO(1,1)\times SO(9)$ transformations (if we allow these to act also on $\lambda_q^{+s}$),  using this symmetry as an identification relation, we can consider $v_{\alpha}^{-q}$ as a kind of homogeneous coordinates of celestial sphere ({\it cf.} (\ref{v-q=S9}))
 \begin{eqnarray}\label{v-q=S9A} \{ v_{\alpha}^{-q}\} = {\bb S}^9
 \; . \end{eqnarray}

One can  see that the  described algebraic  properties of $v_{\alpha}^{-q}$ are the same as that of
spinor moving frame  variables $v_{\alpha}^{-q}(\tau)$ used in the generalized superparticle models of sec. 3. In this section we do not use this name  (neither its shorter version 'spinor frame variables' \cite{Bandos:2017zap})
as it might be confusing  in the context of superalgebra  representations.
Notice however, that the similarity of variables marking states in this section with 1d fields of sec. 3  is not occasional: the quantization of  the superparticle models of sec. 3 should result in the multiplet of quantum states  transforming under representations  which we discuss in this section.

The space of states $|{\cal A}, v_{\alpha q}^{\;-}, \lambda_q^{+s} >$ splits into the sectors
$\{ |{\cal A}, v_{\alpha q}^{\;-}, \lambda_q^{+s} ; r > \}$
with different ranges of the values of index $s$:
$\; s=1,...,r\leq 16$. In the sector with $r=16$ a special role is played by  the states  with
  $\lambda_q^{+s}=\sqrt{2\rho^\#} \delta_q^{\;s}$,
\begin{eqnarray}\label{S0=Sl=I}
  |{\cal A}, v_{\alpha q}^{\;-}, \rho^{\#} > := |{\cal A}, v_{\alpha q}^{\;-}, \sqrt{2\rho^\#} \delta_q^{\;s} ; 16> \; .   \end{eqnarray}
On such states the eigenvalues of the generalized momenta $Z_{ab}$ and $Z_{abcde}$ vanish
and only (super)Poincar\'e generators  are realized nontrivially,
  \begin{eqnarray}\label{cPS0=pGS0}
{\cal P}_{\alpha\beta} \; |{\cal A}, v_{\alpha q}^{\;-}, \rho^{\# }>=\;  2\rho^\# \, v_{\alpha}^{-q}v_{\beta}^{-q} \; |{\cal A}, v_{\alpha q}^{\;-},  \rho^{\# } >\;=\;  p_a\Gamma^a_{\alpha\beta} \; |{\cal A}, v_{\alpha q}^{\;-} , \rho^{\# } >\; ,     \end{eqnarray}
with eigenvalues determined by $ p_a\Gamma^a_{\alpha\beta} =  2\rho^\# \, v_{\alpha}^{-q}v_{\beta}^{-q}$ in terms of  constrained spinors  (\ref{v-q=S9A}) and densities $\rho^{\# }$ ({\it cf. } (\ref{p=u=}), (\ref{v-v-=Gu--})).

\subsection{Some unitary highest weight representations of the M-algebra}

Representations of the supersymmetry generators on the  states
$|{\cal A}, v_{\alpha q}^{\;-}, \lambda_q^{+s} >$ can be characterized by equation
 \begin{eqnarray}\label{Qa=lvG}
 Q_\alpha = \, v_{\alpha}^{-q} \lambda_q^{+s} \; \mathfrak{C}_{s }
 \; , \qquad q=1,...,16\; , \qquad s=1,...,r\leq 16\; ,  \end{eqnarray}
where $\mathfrak{C}_{s }$ are generators of $r$-dimensional Clifford algebra
 \begin{eqnarray}\label{GG=I}
{}\{ \mathfrak{C}_{s } , \mathfrak{C}_{t}\}=2\delta_{st}
 \; , \qquad s,t=1,..., r \; .   \end{eqnarray}
In the context of the above discussion, a more rigorous way is to write
\begin{eqnarray}\label{QaS=lvGS}  Q_\alpha |{\cal A}, v_{\alpha q}^{\;-}, \lambda_q^+ > = \, v_{\alpha}^{-q} \lambda_q^{+s} \; \mathfrak{C}_{s } |{\cal A}, v_{\alpha q}^{\;-}, \lambda_q^+ >\; ,  \end{eqnarray}  where the action of Clifford operator $\mathfrak{C}_{s } $ on the state is still to be defined.
The representation of M-algebra is described  by (\ref{QaS=lvGS}) completed by (\ref{cP=vvll})  which in a more schematic form, similar to (\ref{Qa=lvG}), reads ({\it cf.}  (\ref{pab=rpqvvA}))
\begin{eqnarray}\label{cP=vvll}
{\cal P}_{\alpha\beta} \; =  \lambda_q^{+s} \lambda_p^{+s} \, v_{\alpha}^{-q}v_{\beta}^{-p} \; . \end{eqnarray}

Notice that (\ref{QaS=lvGS}) implies that only $r$($\leq 16$) of 32 supersymmetries are realized nontrivially on the M-algebra representations under consideration. Thus we will be working with short or BPS multiplets of states; all the states of such supermultiplet preserve $32-r$ supersymmetries and only $r$ of the supersymmetry generators mix the different states. If choosing  $r=1$, we would be dealing with preonic multiplets preserving all but one supersymmetries \cite{Bandos:2001pu,Bandos:2003ng}. Here we will be interested mainly  in a more conventional type of multiplets, with $r=16$, all the states of which preserve one half, {\it i.e.} 16 of 32 supersymmetries.

In the case of  even $r$, to construct unitary highest weight representations of the M-algebra, following  the line of \cite{Gunaydin:1998bt} and  using the above type of the eigenstates of generalized momenta, we have to introduce a kind of complex structure and to split the set of $r$ Hermitian generators of Clifford algebra, $ \mathfrak{C}_{s} $, on two conjugate sets of $r/2$ generators, $\mathfrak{B}_{A}$ and $\mathfrak{B}^{\dagger A}$  obeying
\begin{eqnarray}\label{BB+=I}
{}\{ \mathfrak{B}_{A}, \mathfrak{B}_{B}\}=0\; , \qquad {}\{ \mathfrak{B}^{\dagger A}, \mathfrak{B}_{B}\}= \delta_{B}{}^A\; , \qquad \{ \mathfrak{B}^{\dagger A},\mathfrak{B}^{\dagger B}\}=0 \; , \qquad A,B=1,...,\frac r 2 \; . \qquad  \end{eqnarray}
(An explicit form of the  relation between $ \mathfrak{C}_{s} $ and  $\mathfrak{B}_{A}$ and $\mathfrak{B}^{\dagger A}$ will be discussed below).

Then   we can define the highest weight state
 $|\; v_{\alpha}^{-q} , \; \lambda_q^{+s}>$ by
\begin{eqnarray}\label{BS=0}
\mathfrak{B}_{A}|\; v_{\alpha}^{-q} , \; \lambda_q^{+s}> = 0\; , \qquad A=1,...,\frac r 2 \end{eqnarray}
and construct the states of the unitary representation of  the M-algebra by acting on that by (products of) $\mathfrak{B}^{\dagger B}$ operators.

For the case of ${r=16}$, which is of our main interest here,
in such a way we arrive at the representation with 128(=1+28+70+28+1)  bosonic  states
\begin{eqnarray}\label{bose=St}
|\; v_{\alpha}^{-q} , \; \lambda_q^{+s}> \; , \quad
\mathfrak{B}^{\dagger A}\mathfrak{B}^{\dagger B}|\; v_{\alpha}^{-q} , \; \lambda_q^{+s}> \; , \quad \mathfrak{B}^{\dagger A_1}\ldots \mathfrak{B}^{\dagger A_4} |\; v_{\alpha}^{-q} , \; \lambda_q^{+s}>\; ,  \quad \nonumber \\ \mathfrak{B}^{\dagger A_1}\ldots  \mathfrak{B}^{\dagger A_6}|\; v_{\alpha}^{-q} , \; \lambda_q^{+s}> \; , \quad   \mathfrak{B}^{\dagger A_1}\ldots \mathfrak{B}^{\dagger A_8}|\; v_{\alpha}^{-q} , \; \lambda_q^{+s}> \;  \quad \end{eqnarray}  and 128(=8+56+56+8)  fermionic states   \begin{eqnarray}\label{fermi=St}
\mathfrak{B}^{\dagger A}|\; v_{\alpha}^{-q} , \; \lambda_q^{+s}> \; , \quad
\mathfrak{B}^{\dagger A}\mathfrak{B}^{\dagger B}\mathfrak{B}^{\dagger C}|\; v_{\alpha}^{-q} , \; \lambda_q^{+s}> \; , \quad \mathfrak{B}^{\dagger A_1}\ldots \mathfrak{B}^{\dagger A_5}|\; v_{\alpha}^{-q} , \; \lambda_q^{+s}> \; , \quad \nonumber \\  \mathfrak{B}^{\dagger A_1}\ldots \mathfrak{B}^{\dagger A_7}|\; v_{\alpha}^{-q} , \; \lambda_q^{+s}> \; .  \quad   \end{eqnarray}

We claim that, when $\lambda_q^{+s}=\sqrt{2\rho^\#} \delta_q^{\;s}$, $s=1,...,16$, the states of the above described unitary highest weight representation of the M-algebra can be identified with degrees of freedom of the eleven-dimensional  supergravity \cite{Cremmer:1978km}. Then, in the case of generic $\lambda_q^{+s}$,  the states (\ref{bose=St}), (\ref{fermi=St}) can be associated to the fields of 11D supergravity multiplet depending, besides 11-vector coordinate or momenta, on a set of 135 additional variables. These latter can be described by
\begin{eqnarray}\label{varphi=} & \varphi_q^s= \frac {\lambda_q^{+s} }{\sqrt{\lambda_p^{+t}\lambda_p^{+t}}} \;  \qquad \end{eqnarray} defined modulo $O(16)$ transformations:
\begin{eqnarray}\label{varphi=equiv}
\varphi_q^s \approx \varphi_q^t {\cal O}_t{}^s\; , \qquad {\cal O}{\cal O}^T=I_{16\times 16}\; . \qquad \end{eqnarray}

\subsection{Unitary highest weight representations with $r=16$ and 11D supergravity}

One of the way to see the above claimed relation of $r=16$ unitary highest weight representation of the M-algebra with 11D supergravity starts form a seemingly different representation of the generators
on the set of
128 bosonic states, 44 of which are enumerated  by the symmetric traceless pair of SO(9) vector indices, ${\cal A}=IJ=((IJ))$, and remaining 84 - by the set of three antisymmetric SO(9) vector indices, ${\cal A}=IJK=[IJK]$, and 128 fermionic states enumerated by the gamma-traceless set of 9-vector and SO(9) spinor indices,
${\cal A}=Is$,
 \begin{eqnarray}\label{statesB}
128&=84+44&: \qquad |IJK, \;v_{\alpha q}^{\;-}, \lambda_q^{+s} > =|[IJK], \;v_{\alpha q}^{\;-}, \lambda_q^{+s} > \; , \qquad \nonumber \\ && |IJ,\;  v_{\alpha q}^{\;-}, \lambda_q^{+s} > = |JI,\;  v_{\alpha q}^{\;-}, \lambda_q^{+s} >
\; , \qquad |II, \; ...  > \equiv 0\; ,\qquad  \\ \label{statesF}
128&=144-16&: \qquad |Is, \;  v_{\alpha q}^{\;-}, \lambda_q^{+s'} > \; , \qquad \gamma^I_{st}|It, \; ... >  \equiv 0\; , \qquad \end{eqnarray}
where $\gamma^I_{st}$ are 9d Dirac matrices ($s,t=1,...,16$).
This representation is described by Eqs. (\ref{cPS=vvllS}) and  (\ref{QaS=lvGS}) with the following action of 16 hermitian  Clifford generators $\mathfrak{C}_s$ on the above states
\begin{eqnarray}
\label{repr-16b}
 \mathfrak{C}_s |IJK, v_{\alpha q}^{\;-}, \lambda_q^{+s'} >  =   \gamma^{IJ}_{st} |Kt, v_{\alpha q}^{\;-}, \lambda_q^{+s'} > + \gamma^{KI}_{st}
|Jt, v_{\alpha q}^{\;-}, \lambda_q^{+s'} > + \gamma^{JK}_{st}|It, v_{\alpha q}^{\;-}, \lambda_q^{+s'} >  , \qquad \nonumber \\ \mathfrak{C}_s \; |IJ, v_{\alpha q}^{\;-}, \lambda_q^{+s'} >   = \gamma^I_{st} |Jt, v_{\alpha q}^{\;-}, \lambda_q^{+s'} > +
\gamma^J_{st}
|It, v_{\alpha q}^{\;-}, \lambda_q^{+s'} >  , \qquad  \\
\label{repr-16f}
\mathfrak{C}_s \; |It, v_{\alpha q}^{\;-}, \lambda_q^{+s'} > = \gamma^J_{st}|IJ, v_{\alpha q}^{\;-}, \lambda_q^{+s'} > + {1\over 3!}
\left(\gamma^{IJKL}_{st} - 6 \delta^{I[J} \gamma^{KL]}_{st}\right)
|{JKL},v_{\alpha q}^{\;-}, \lambda_q^{+s'} >. \qquad
\end{eqnarray}

Such a representation of the M-algebra with $\lambda_q^{+s}=\sqrt{\rho^\#}\delta_q^{\;s}$
was obtained in \cite{Bandos:2007wm} as a result of covariant quantization of the 11D Brink-Schwarz superparticle in its spinor moving frame formulation. Essentially the same representation had been obtained in the light-cone quantization \cite{Green:1999by}.
The bosonic and fermionic fields corresponding to the basic vectors of these representation
\begin{eqnarray}\label{states=wfbs}
A_{IJK}(v_{\alpha q}^{\;-}, \rho^\# )\leftrightarrow  |IJK, \;v_{\alpha q}^{\;-}, \rho^\# > \; , \qquad h_{IJ}(v_{\alpha q}^{\;-}, \rho^\# ) \leftrightarrow  |IJ,\;v_{\alpha q}^{\;-}, \rho^\# >
\; , \qquad \nonumber \\ A_{IJK}=A_{[IJK]}\; , \qquad h_{IJ}=h_{JI}\; , \quad h_{II}= 0\; ,\qquad  \\ \label{states=wff}
\Psi_{Iq} (v_{\alpha q}^{\;-}, \rho^\# )  \leftrightarrow |Iq, \;  v_{\alpha q}^{\;-}, \rho^\# > \; , \qquad \gamma^I_{qp}\Psi_{Ip} \equiv 0\; , \qquad \end{eqnarray}
describe the on-shell degrees of freedom of the  11D supergravity. Indeed, using  (\ref{v-q-q=}), (\ref{u--u--=0}), (\ref{v-v+=GuI}) and (\ref{v-v-=0}), one can easily check that
\begin{eqnarray}\label{Fabcd=}
F_{abcd}= \rho^\# u_{[a}^=u_{b}^Iu_{c}^Ju_{d]}^K A_{IJK}\; , \qquad
R_{abcd}= (\rho^\#)^2 u_{[a}^=u_{b]}^Iu_{[c}^=u_{d]}^J h_{IJ}\; , \qquad \\  \label{Psiabal=} \Psi_{a}^\alpha =
u_{a}^I v_q^{-\alpha}\Psi_{Iq}\;   \qquad \end{eqnarray}
  solve the linearized field equations of  11D SUGRA in the momentum representation,
\begin{eqnarray}\label{Eqs=lin}p^aF_{abcd}=0\; , \qquad R_{ab}{}^{cb}=0\; , \qquad  p_a\Gamma^a_{\alpha\beta}\Psi_{b}^\beta =0\; , \quad  p^a\Psi_{a}^\alpha =0\; , \quad \Gamma^a_{\alpha\beta}\Psi_{a}^\beta =0\; , \quad  \end{eqnarray}
\footnote{It is not difficult to check that the Rarita-Schwinger equation, $ \Gamma^{abc}_{\alpha\beta}p_b\Psi_{c}^\beta =0$, follows from last three equations in (\ref{Eqs=lin}).} provided the light-like momentum is related to the constrained spinors by
 \begin{eqnarray}\label{pa=rpvv}p_a= \rho^{\#}  u_a^=  \qquad  \Leftrightarrow \qquad  p_a\Gamma^a_{\alpha\beta}= 2 \rho^{\#} \, v_{\alpha}^{-q}v_{\beta}^{-q}\; . \end{eqnarray}

The  M-algebra representations with generic $\lambda_q^{+s}$ are characterized by nonvanishing eigenvalues of tensorial momentum generators $Z_{ab}$ and $Z_{abcde}$. The basic states of such a representation can be represented by on-shell fields of 11D supergravity depending on additional variables,
\begin{eqnarray}\label{states=wfbs+}
A_{IJK}(v_{\alpha q}^{\;-}, \rho^\# ; \varphi_q{}^s)\leftrightarrow  |IJK, \;v_{\alpha q}^{\;-}, \lambda_q^{+s}> \; , \qquad h_{IJ}(v_{\alpha q}^{\;-}, \rho^\# , \varphi_q{}^s) \leftrightarrow  |IJ,\;v_{\alpha q}^{\;-}, \lambda_q^{+s} >
\; , \qquad \nonumber \\ A_{IJK}=A_{[IJK]}\; , \qquad h_{IJ}=h_{JI}\; , \quad h_{II}= 0\; ,\qquad  \\ \label{states=wff+}
\Psi_{Is} (v_{\alpha q}^{\;-}, \rho^\# , \varphi_q{}^s)  \leftrightarrow|Is, \;  v_{\alpha q}^{\;-}, \lambda_q^{+s}> \; , \qquad \gamma^I_{st}\Psi_{It} \equiv 0\; , \qquad \end{eqnarray}
where $\rho^\# = \lambda_q^{+s}\lambda_q^{+s}$ and $\varphi_q{}^s$ is defined in (\ref{varphi=}), (\ref{varphi=equiv}).

To show that the representation of M-algebra which is  described by (\ref{Qa=lvG}), (\ref{cP=vvll}), (\ref{repr-16b}), (\ref{repr-16f}) can be identified with the highest weight representation (\ref{bose=St}), (\ref{fermi=St}), we have to introduce a complex structure which  breaks the natural SO(9)($\subset SO(16)$) symmetry of our construction down to
Spin(7)($\subset SU(8)$). This is achieved by introducing a complex null vector $U_I$, obeying
\begin{eqnarray}\label{U2=0}
U_IU_I=0\; , \qquad \bar{U}_I\bar{U}_I=0
\; , \qquad U_I\bar{U}_I=2\; , \qquad \bar{U}_I=({U}_I)^*\; , \qquad \end{eqnarray}
 and the rectangular 16$\times$ 8  complex conjugate matrices $w_{s}{}^A$ and $\bar{w}_{As}=(w_{s}{}^A)^*$ obeying
\begin{eqnarray}
\label{bww=1}
&& \bar{w}_{sB}w_{s}{}^A =\delta_B{}^A\; , \qquad w_{s}{}^A w_{s}{}^B =0 \; , \qquad \bar{w}_{sA} \bar{w}_{sB} =0\;  \qquad
\end{eqnarray}
and
\begin{eqnarray}\label{U/=wcUw}
{U}{}_I\gamma^I_{st}= 2\bar{w}_{sA} \bar{w}_{tA}\; , \qquad
{}
 \bar{U}{}_I\gamma^I_{st}= 2w_s^{\;A} w_t^{\; A}\; \qquad
\end{eqnarray}
involving   9d gamma matrices $\gamma^I_{st}$.

Then the  relation of the complex and hermitian generators of Clifford algebra is
\begin{eqnarray}\label{BA=wC}
\mathfrak{B}_A =  \bar{w}_{sA} \mathfrak{C}_s \; , \qquad  \mathfrak{B}^{\dagger A} = w_s^A \mathfrak{C}_s \; , \qquad
 A=1,...,8\; , \qquad s=1,...,16\; ,
\end{eqnarray}
and the highest weight vector is defined by
\begin{eqnarray}\label{hweight=}
  | v_{\alpha q}^{\;-}, \lambda_q^{+s} > := U_I U_J |IJ, v_{\alpha q}^{\;-}, \lambda_q^{+s} >  \; .   \end{eqnarray}
Indeed, using (\ref{repr-16b}), (\ref{bww=1}) and (\ref{U/=wcUw}) it is easy to check that this vector obeys (\ref{BS=0}) with
$\mathfrak{B}_A$ from (\ref{BA=wC}).

This completes the proof of the equivalence of highest weight representation  (\ref{bose=St}), (\ref{fermi=St}) and  the representation defined by (\ref{Qa=lvG}), (\ref{repr-16b}), (\ref{repr-16f}), which contains 11D supergravity by construction \cite{Green:1999by,Bandos:2007wm}.

\subsection{On moduli space of (complex structures defining) the highest weight representations}

Some comments concerning new objects (\ref{bww=1}) and (\ref{U2=0}) defining the complex structures characterizing the above discussed  highest weight representations might  be useful. As a simplest possibility, one can think about fixed null-vector $U_I= \delta_I^8+i\delta_I^9$ and chose $\bar{w}_{sA}$ to be an arbitrary factorization of
complex nilpotent (rank 8) matrix $\gamma^8+i\gamma^9$,
$$ (\gamma^8+i\gamma^9)_{st}=2 \bar{w}_{sA}\bar{w}_{tA}\; ,\qquad (\gamma^8-i\gamma^9)_{st}=2 w_s^Aw_t^A\; ,\qquad (I-i\gamma^8\gamma^9)_{st}=2 \bar{w}_{sA}w_s^A\; .\qquad   $$

A generic complex structure on 16-dimensional Clifford algebra  can be defined by $(\bar{w}_{sA}, w_s^A)$ formed from the columns of
Spin(9) valued matrix $w_s^{(t)}$. Then the light-like vector $U_I$ is formed from the columns of the SO(9) valued matrix
\begin{eqnarray}\label{UinSO9}
&  U_I^{(J)}= \left(U_I{}^{\check{J}}, U_I{}^{(8)}, U_I{}^{(9)}\right)= \left(U_I{}^{\check{J}}, \frac 1 2 \left( U_I+ \bar{U}_I\right), \frac 1 {2i} \left( U_I- \bar{U}_I \right)\right) \; \in \; SO(9)
 \;  \qquad
\end{eqnarray}
related to $w_s^{(t)}\in Spin(9)$ by
$U_I^{(J)} \gamma^I_{qp} = w_q^{(q')} \gamma^{(J)}_{(q')(p')} w_p^{(p')}$ (see \cite{Bandos:2017zap} for details).

Thus, starting from  one 11D supergravity--related representation (\ref{states=wfbs+}),
we can construct a family of unitary highest weight representations, elements  of which are characterized
by complex structures  described by $(\bar{w}_{sA}, w_s^A)\in Spin(9)$
defined modulo $Spin(7)\times U(1)$ transformations, i.e. parametrizing the coset $\frac {Spin(9)}{ Spin(7)\times U(1)}$. In this sense we can say that the moduli space of (complex structures defining) the
highest weight representations constructed on the basis of one 11D supergravity-related representation is
\begin{eqnarray}\label{Mcopl}
{\cal M}_{hw}= \frac {Spin(9)}{ Spin(7)\times U(1)}
 \; . \qquad
\end{eqnarray}

\bigskip

It is natural to expect  that the above discussed unitary highest weight representations  as well as the equivalent, explicitly 11D  SUGRA-related representations (\ref{repr-16b}), (\ref{repr-16f}), can be obtained as a result of  quantization of generalized superparticle models described by  (\ref{S=lvlvdX}) with (\ref{rpq=lsls}) and $r=16$.
Such a quantization and the analysis of quantum state spectra are still to be done and we hope to address this problem in a forthcoming publication.

\section{Conclusions and discussion }

\label{Conclusion}

In this paper we conjectured the existence of hypothetical underlying exceptional field theory, which we
abbreviated as 11D EFT or uEFT, defined in the maximal tensorial central charge superspace
\begin{eqnarray}\label{S528-32}
\Sigma^{(528|32)} = \; \{x^a, y^ {ab}, y^{abcde}, \theta^\alpha \}\; , \quad a,b,c,d,e =0,1,..., 9,10 \; , \quad \alpha=1,...,32\; , \end{eqnarray}
which is the group manifold associated to the M-theory superalgebra (\ref{susyA11D}).
We have presented some arguments in favor of this conjecture, based on the hypothesis that the additional coordinates of all the $E_{n(n)}$ EFTs with $n\leq 8$ should  be related to the  maximally extended $d=11-n$ dimensional supersymmetry algebra, and have proposed the candidate section conditions for this  hypothetical uEFT
\begin{eqnarray}\label{11Dsec=d1}
&&  \partial_{[a} \otimes \partial_ {bc]} +  \partial_{[bc} \otimes \partial_{a]}=0\; ,   \qquad
\partial_{[a} \otimes \partial_{bcdef]} + \partial_{[bcdef} \otimes  \partial_{a]} =0,  \\ \label{11Dsec=d2} && \partial_{[ab} \otimes \partial_{bc]} =0\; , \qquad
\partial_{[ab} \otimes \partial_ {cdefg]} +  \partial_{[cdefg} \otimes \partial_{ab]}=0\ \; . \qquad
\end{eqnarray}
To check that these section condition are reasonable, i.e. that their general solution is not trivial,  we have discussed a series of superparticle models in $\Sigma^{(528|32)}$  and show that they produce quite generic solutions  of (the classical counterparts of) these hypothetical section conditions as constraints on their generalized momenta. Of course, the next question was what is the physical meaning of these superparticle models. To address it  we have presented some arguments that these superparticle models  should produce free massless 11D higher spin field theories as  their quantum state spectrum.

We have also discussed some unitary highest weight representations of M-theory superalgebra
and the embedding of 11D supergravity in such representations. Besides giving an additional argument in favour of relevance of
$\Sigma^{(528|32)}$ superspace, and of our uEFT hypothesis, this provides us with a hint about how the quantum state spectrum of some of the generalized superparticle models might look like. Namely, the representations preserving   one half (16 of 32) supersymmetries, presumably related with some special class of  $\Sigma^{(528|32)}$ superparticle models, can be equivalently described by on-shell fields of 11D supergravity depending on additional variables.

By passing, we have also argued that 10D ($m=16$) and 6D ($m=8$) counterparts of these superparticle models, defined in  $\Sigma^{\left(\frac {m(m+1)} 2 | m\right)} $  superspaces with $m=2(D-2)$, provides a classical mechanics description of free non-conformal massless higher spin theories in D=10 and D=6. These are of interest because the most interesting massless fields, like D-dimensional graviton and photon, are not conformal in $D\not=4$.

The above observations suggest that the hypothetical underlying uEFT or 11D EFT  should contain the 11D higher spin theory as an important sector. Interestingly enough,  11D Higher spin fields were recently considered as a probably necessary ingredients of completion of  11D supergravity till a $E_{10}$ or even  $E_{11}$ invariant theories \cite{Henneaux-Talk}.

Even before, the relation of $E_{11}$ with higher spin theories was discussed in
\cite{West:2007mh} where the action of \cite{Bandos:1998vz} supplemented by the condition of light-likeness of the bilinear $\lambda \Gamma_a\lambda$, is proposed as a candidate for low level (three level) approximation of  a hypothetical point particle model based on a non-linear realization of $E_{11}\subset\!\!\!\!\!\!\times l_1$. Here
$l_1$ is the fundamental representation associated with the 'far end' of the $E_{11}$ Dynkin diagram, usually called 'node 1' (see e.g. \cite{West:2010rv}), the (infinite) set generators of which  contains,  at lowest level, $P_a$, $Z^{ab}$, $Z^{abcde}$, which can be identified with the bosonic generators of M-algebra (\ref{susyA11D}).

Such a three level approximation to  hypothetical $E_{11}$ superparticle  can be identified with the 'preonic' representative (\ref{S=lvlvdX}) of the family of the action of uEFT superparticles (\ref{S=lplq}), (\ref{S=rpq}). It would be interesting to understand a possible role of other representatives of this  family in an $E_{11}$ perspective.

Thus 11D higher spin theories are probably common ingredients of the uEFT and of the $E_{11}$ and $E_{10}$ theories. On the other hand, these are not identical, but rather complementary. As it is seen in Table 1,
$E_{n(n)}$ EFTs, making manifest the the rigid $E_{n(n)}$ duality symmetry, keep only the
$SO(1, 10-n)$ subgroup of the 11D $SO(1,10)$ Lorentz group manifest. A part of Lorentz symmetry is sailed for an increase of the manifest rigid symmetry \footnote{As $E_{n(n)}$ EFT is an extension of  $d$ dimensional supergravity, the Lorentz $SO(1,d-1)$ symmetry is one of its manifest gauge symmetries while
$E_{n(n)}$ is the rigid symmetry. Although our discussion here used the models in flat superspaces where the  SO(1,D-1) is not local, our terminology in this  section is borrowed from the complete description of EFTs.}.  In this consequence $E_{10}$ and $E_{11}$ clearly correspond to $d=1$ and $d=0$ (!), thus not leaving any part of the  $SO(1,10)$ Lorentz symmetry  manifest, while our
uEFT correspond to D=11 and $n=0$, thus leaving all the Lorentz symmetry but no rigid symmetry manifest.
Together with the Lorentz symmetry this hypothetical underlying theory should possess manifest supersymmetry the generators of which form the M-theory superalgebra, the  maximally enlarged supersymmetry algebra in 11D.
The uEFT is defined in the 11D superspace enlarged by coordinate conjugate to the  tensorial central charge generators of this superalgebra, (\ref{S528-32}), which provides the resource for all the additional coordinates of the lower $d$/higher $n$ EFTs with $n<9$.

As a potentially suggestive speculation, let us  to try to incorporate the hypothetical $E_{10}$ and $E_{11}$ theories and uEFT in the Table 1. Clearly, the first two should be put on the top and the last - at the bottom of the Table 1, see Table 2 below.

If doing just this, we would have the holes at $n=9$ ($d=2$) and $n=1$ ($d=10$). The first of this positions should clearly correspond to some hypothetical $E_{9}$ EFT having manifest symmetry under the
infinite dimensional Katz-Moody algebra $E_9$ \cite{Julia:1982gx}.
The second, $n=1$ hole we have filled in  Table 2 by putting the  doubled field theory, DFT, in this line.
The main motivation is that this case clearly correspond to d=10, so that its association with $n=1$ comes just from $n=11-d$. The increasing of the number of additional coordinates till $10$, in contrast to their decreasing from 248 to 3 when $d$ increased from 3 to 9, might mark the  change of tendency which is then continuing on the stage of passing to 11D uEFT, which is defined in (super)space with 517=528-11 additional coordinates.  Our hypothesis does not associates the T-duality group $O(10,10)$ with a hypothetical  $E_{1(1)}$ as far as the 10d Lorentz group $SO(1,9)$ is a subgroup of $O(10,10)$. Rather $E_{1(1)}$ should be associated with the generators of the  coset $O(10,10)/SO(1,9)$ so that, although a big number of duality symmetries is present, they do not form a closed algebraic structure themselves, but together with Lorentz group only.

\begin{center}
    \begin{tabular}{ | l | l | l | l | p{5cm} |}
    \hline
    $E_{n(n)}$ of the EFT & n & d=11-n & $N_n$=$\#$ of $y^\Sigma$   & Section condition \\ \hline  & & & &  \\
    ${\bf ? }\!\!\!/$ $E_{11}$ & 11 & d=0 & {\LARGE $\infty$}    & ${\bf ? }\!\!\!/$ see recent \cite{Bossard:2017wxl}\\
    ? $E_{10}$ & 10 & d=1 & {\Large $\infty$}     & {\bf ? }
    \\
     $E_{9}$ & 9 & d=2 & $\infty$    & see \cite{talk-Martin-2017} \\
    \hline  & & & &  \\
    $E_{8(8)}$ & 8 & d=3 & 248   & $Y_{\Lambda\Xi}{}^{{\Sigma}{\Pi}}\partial_{{\Sigma}} \otimes \partial_{{\Pi}} =0$,  \cite{Hohm:2014fxa}\\   $E_{7(7)}$ & 7 & d=4 & 56  & $t_{G}^{{\Sigma}{\Pi}}\partial_{{\Sigma}} \otimes \partial_{{\Pi}} =0$, \cite{Hohm:2013uia} \\
    $E_{6(6)}$ & 6 & d=5 & 27  &  $d^{{\Lambda}{\Sigma}{\Pi}}\partial_{{\Sigma}} \otimes \partial_{{\Pi}} =0$  \cite{Hohm:2013pua}  \\
    $E_{5(5)}=SO(5,5)$ & 5 & d=6 & 16  & $\gamma_{I}^{{\Sigma}{\Pi}}\partial_{{\Sigma}} \otimes \partial_{{\Pi}} =0$ \cite{Abzalov:2015ega} \\
     $E_{4(4)}=SL(5)$ & 4 & d=7 & 10 ($y^{\tilde{\mathfrak{a}}\tilde{\mathfrak{b}}}=y^{[\tilde{\mathfrak{a}}\tilde{\mathfrak{b}}]}$) & $\partial_{[\tilde{\mathfrak{a}}\tilde{\mathfrak{b}}}\otimes \partial_{\tilde{\mathfrak{c}}\tilde{\mathfrak{d}}]}=0$ \cite{Berman:2011cg,Blair:2013gqa} \\
    $E_{3(3)}=SL(3)\times SL(2)$ & 3 & d=8 & 6 $(y^{\alpha i})$  & $\epsilon^{ijk}\epsilon^{\alpha\beta} \partial_{{\alpha} i} \otimes \partial_{{\beta} j} =0$  \cite{Hohm:2015xna}
    \\ $E_{2(2)}=SL(2)\times {\bb R}^+$ & 2 & d=9 & 3 $(y^\alpha, z)$ & $ \partial_{z} \otimes \partial_{\alpha}+ \partial_{\alpha} \otimes \partial_{z}  =0$ \cite{Berman:2015rcc} \\
     \hline  & & & &  \\
     DFT: $SO(10,10)$ & 1?  & d=10 & 10 ($\tilde{x}_\mu$) & $\partial_\mu \otimes   \tilde{\partial}^\mu  + \tilde{\partial}^\mu \otimes \partial_\mu =0$ \cite{Siegel:1993xq,Hull:2004in}  \\ &&&& \\
     {\bf uEFT}: only  $SO(1,10)$ & 0?  & d=11 & 528 ($y^{ab}, y^{abcde})$ & $\partial_{[a} \otimes \partial_ {bc]} +  \partial_{[bc} \otimes \partial_{a]}=0$,  \\ is manifest & &&&
$\partial_{[a} \otimes \partial_{bcdef]} + \partial_{[bcdef} \otimes  \partial_{a]} =0$,  \\ &&&& $\partial_{[ab} \otimes \partial_{cd]} =0$, {\it etc.} (see (\ref{11Dsec=d2})) \\
    \hline
    \end{tabular}

\medskip

Table 2. Known and hypothetical EFTs. Conjectured places of the underlying  EFT and of the hypothetical EFTs for infinite dimensional groups in Table 1.
\end{center}

\medskip

The above discussion suggests the origin of doubled spacetime coordinate of DFT in the uEFT superspace $\Sigma^{(528|32)}$, and furthermore,  that our 11D uEFT provides a unification of DFT with
$n=2,...,8$ $E_{n(n)}$ exceptional field theories.

We hope that our underlying 11D exceptional field theory (11D uEFT) conjecture  will be useful for deeper understanding of dualities and of the structure of/beyond the  M-theory. It will be interesting to elaborate further on its interrelation/complementary with the $E_{11}$ and $E_{10}$ proposal. One of the important directions of  further development of our approach is related to the quantization of the
family of the generalized superparticle models (\ref{S=lplq}). In particular this should provide  us with a description of towers of massless non-conformal higher spin theories in D=11, the possible fundamental role of which was also a  subject of thinking in the Higher Spin community \cite{Vasiliev=priv-com}.

\bigskip

\section*{Acknowledgements}
This work has been supported in part by the Spanish Ministry of Economy, Industry and Competitiveness  grants FPA 2012-35043-C02-01 and FPA 2015-66793-P, partially financed  with FEDER funds,  by the Basque Government research group grant
IT-979-10,
and by the Basque Country University program UFI 11/55.
The author is thankful to Dima Sorokin, Mikhail Vasiliev, to David Berman, Martin Cederwall, Jakob Palmkvist, Jeong-Hyuck Park, Henning Samtleben for useful discussions,
to the Munich Institute for Astro- and Particle Physics (MIAPP) of the DFG cluster of excellence "Origin and Structure of the Universe" (Munich, Germany), to the Galileo Galilei Institute for Theoretical Physics and the INFN (Florence, Italy) and to BIRS (Banff International Research Station
for Mathematical Innovation and Discovery, Banff, Alberta, Canada)  for the hospitality and partial support of his visits at certain stages of this work.

\section*{Notice added.} After this paper have been completed, the
$E_{11}$ EFT was constructed  in \cite{Bossard:2017wxl}. It is based (and develops)
the theory non-linear realization of $E_{11}\otimes {\ell}_1$ \cite{West:2001as}--
\cite{West:2010rv}, \cite{West:2014eza}--\cite{Tumanov:2016dxc}, in which the fields depend on infinitely many coordinates (of ${\ell}_1$),  but also presents a set of section conditions, which results in the dependence of fields on the finite number of coordinates. (This has allowed us to replace
$?\mapsto {\bf ? }\!\!\!/$ in the first line of Table 2).
An even more  extended scheme based on
infinite--dimensional  tensor hierarchy superalgebra is also proposed in \cite{Bossard:2017wxl}.

\end{document}